\documentclass{ieeetmlcn}
\AtBeginDocument{\definecolor{tmlcncolor}{cmyk}{0.93,0.59,0.15,0.02}\definecolor{NavyBlue}{RGB}{0,86,125}}
\def\OJlogo{\vspace{-4pt}\includegraphics[height=18pt]{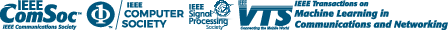}}
\def\seclogo{\vspace{10pt}\includegraphics[height=18pt]{new_logo}}
\def\authorrefmark#1{\ensuremath{^{\textbf{#1}}}}
 
\usepackage{graphicx}
\usepackage{epstopdf}
\usepackage{float}
\usepackage{caption}
\usepackage{subcaption} 
\usepackage[ruled,vlined]{algorithm2e} 
\usepackage{amsmath} 
\usepackage{amsthm,amssymb}
\usepackage{amsfonts}
\usepackage{bm}
\usepackage{dsfont}
\usepackage{xcolor}
\usepackage{url}
\usepackage{amssymb}
\usepackage{multirow}
\usepackage{graphicx}

\newcommand{\cmark}{\checkmark}

\usepackage{stmaryrd}

\usepackage{array,booktabs,multirow,makecell}
\usepackage{tabularx}
\usepackage{colortbl}

\usepackage{color}

\allowdisplaybreaks[4]

\definecolor{headercolor}{RGB}{52, 101, 164}
\definecolor{rowcolor1}{RGB}{242, 246, 252}
\definecolor{rowcolor2}{RGB}{255, 255, 255}

\newcommand{\dis}{:=}
\newtheorem{theorem}{Theorem}

\newtheorem{lemma}{Lemma}

\newtheorem{proposition}{Proposition}

\newtheorem{property}{Property}

\newtheorem{remark}{Remark}

\newtheorem{claim}{Claim}

\newcolumntype{A}{>{\hsize=0.82\hsize\raggedright\arraybackslash}X}
\newcolumntype{B}{>{\hsize=0.6\hsize\centering\arraybackslash}X}
\newcolumntype{C}{>{\hsize=1.7\hsize\centering\arraybackslash}X}
\newcolumntype{D}{>{\hsize=0.75\hsize\centering\arraybackslash}X}

\usepackage{xspace}  % 添加用于处理空格的包
\usepackage{mathtools}  % 增强数学功能
\usepackage{dblfloatfix}

\begin{document}

\receiveddate{XX Month, XXXX}
\reviseddate{XX Month, XXXX}
\accepteddate{XX Month, XXXX}
\publisheddate{XX Month, XXXX}
\currentdate{XX Month, XXXX}
\doiinfo{TMLCN.2022.1234567}

\markboth{}{Author {et al.}}

\title{MAGRPO: Accelerated MARL Training for Fluid Antenna-Assisted Wireless Network Optimization}

\author{Wanzhe Wang\authorrefmark{1}, Yanfei Su\authorrefmark{1}, Tong Zhang\authorrefmark{2}, Hao Xu\authorrefmark{3}, Shuai Wang\authorrefmark{4}, and Rui Wang\authorrefmark{5}}
\affil{Harbin Institute of Technology, Shenzhen, China}
\affil{Guangdong Provincial Key Laboratory of Aerospace Communication and Networking Technology, Harbin Institute of Technology, Shenzhen, China}
\affil{School of Information Science and Engineering, Southeast University, Nanjing, China}
\affil{Shenzhen Institutes of Advanced Technology, Chinese Academy of Sciences, Shenzhen, China}
\affil{Southern University of Science and Technology, Shenzhen, China}
\corresp{Corresponding author: Tong Zhang (email: tongzhang@hit.edu.cn).}
\authornote{Wanzhe Wang and Yanfei Su contributed equally to this work. Code is available at https://github.com/yanfeisu/MAGRPO.git}

\begin{abstract}
Fluid antenna systems (FASs) improve wireless links by repositioning antenna elements to exploit favorable spatial channel variations. Jointly optimizing fluid antenna (FA) positions, beamforming, and transmit power in a multi-cell network is challenging because the problem is non-convex and each base station has only local information during decentralized execution. However, the representative multi-agent reinforcement learning (MARL) algorithms, namely the on-policy multi-agent proximal policy optimization (MAPPO) and the off-policy multi-agent twin delayed deep deterministic policy gradient (MATD3), suffer from excessively long training times. To address this challenge, we formulate the problem as a decentralized partially observable Markov decision process (Dec-POMDP) and propose multi-agent group relative policy optimization (MAGRPO) under centralized training with decentralized execution. MAGRPO constructs relative advantages from groups of joint trajectories, thereby eliminating the centralized critic and generalized advantage estimation used by MAPPO; under parameter sharing, this critic-free design reduces the per-step computational complexity by approximately half. Simulations show that joint FA optimization provides several-fold sum-rate gains over fixed-position configurations. Across the evaluated antenna settings, MAGRPO achieves test sum rates higher than those of MATD3 and comparable to or slightly higher than those of MAPPO. Meanwhile, it reduces the training time by $20\%$--$23\%$ compared with MAPPO and by about $35\%$ compared with MATD3.
\end{abstract}

\begin{IEEEkeywords}
Decentralized partially observable Markov decision process,
fluid antenna system,
multi-agent group relative policy optimization,
multi-agent proximal policy optimization,
multi-agent reinforcement learning.
\end{IEEEkeywords}

\maketitle

\section{Introduction}

\IEEEPARstart{M}{odern} wireless communications have evolved through generational shifts, each marking a significant leap forward. The ongoing transition from the fifth generation (5G) to the sixth generation (6G) is expected to provide ultra-high data rates, massive connectivity, ultra-low latency, and high energy efficiency, which impose stringent requirements on the physical-layer design \cite{MIMO-1,MIMO-2,MIMO-3}. Multiple-input multiple-output (MIMO) is a cornerstone technology of modern wireless systems. Distinguished by its multiplexing and diversity gains, MIMO enables the simultaneous transmission of multiple data streams over the spatial medium while enhancing robustness against fading through multiple independent signal paths, thereby substantially improving the spectral efficiency. Despite these advantages, MIMO performance remains highly dependent on channel conditions because fixed antenna positions cannot adapt to favorable conditions, causing significant performance degradation. To address this limitation, the fluid antenna system (FAS) has emerged as a promising technology that rapidly adjusts the antenna positions. By doing so, FAS actively reshapes the MIMO channel response, converting poor channel conditions into favorable ones and thereby overcoming the limitation of fixed-position MIMO systems \cite{FAS-Kit-1}-\cite{FluidAntenna_Wu2025}. In multi-cell FAS-assisted networks, jointly optimizing the antenna positions, beamforming, and transmit power is nevertheless challenging because the resulting problem is non-convex and each base station (BS) has only local information, which motivates the development of efficient learning-based solutions.

\begin{table*}[t]
\centering
\caption{Comparison of This Work and Related Studies on FA-Assisted Wireless Networks}
\label{tab:related_work}
\footnotesize
\setlength{\tabcolsep}{8pt}
\renewcommand{\arraystretch}{1.0}

\begin{tabular}{@{}c|c|c|c|c|c|c|c|c|c@{}}
\hline
\multirow{2}{*}{\textbf{Work}}
& \multicolumn{3}{c|}{\textbf{Antenna Configuration}}
& \multicolumn{3}{c|}{\textbf{Optimization Variables}}
& \multicolumn{2}{c|}{\textbf{Algorithmic Design}}
& \multirow{2}{*}{\textbf{Analysis}} \\
\cline{2-9}
& SISO
& MISO
& MIMO
& FAS Position
& Beamforming
& Transmit Power
& Numerical Methods
& DRL
& \\
\hline

\cite{wei2025movable} &        & \cmark &        & \cmark & \cmark & \cmark & \cmark &        &        \\
\hline
\cite{zhu2025joint} &        &        & \cmark & \cmark & \cmark &        & \cmark &        &        \\
\hline
\cite{Zhao2026FA} &        & \cmark &        & \cmark & \cmark & \cmark & \cmark &        &        \\
\hline
\cite{mao2025joint} &  \cmark &      &        & \cmark &        &        & \cmark &        &        \\
\hline
\cite{wu2026fluid} &        &  \cmark  &     & \cmark & \cmark &        & \cmark &        & \cmark    \\
\hline
\cite{Ghadi2025} & \cmark &        &        & \cmark &        &        &        &        & \cmark \\
\hline
\cite{Li2026} &        &  \cmark &     & \cmark & \cmark & \cmark & \cmark &        &        \\
\hline
\cite{Li2025MATD3} &        & \cmark &        & \cmark & \cmark &        &        & \cmark &        \\
\hline
\cite{su2024cd} & \cmark &        &        &        &        & \cmark &        & \cmark &        \\
\hline
\rowcolor{rowcolor1}
\textbf{Ours} & \cmark & \cmark &        & \cmark & \cmark & \cmark &        & \cmark &        \\
\hline

\end{tabular}%
\end{table*}

\subsection{Related Work}
Early research focused on point-to-point systems. Specifically, \cite{efrem2022transmit} addressed the transmit and receive FA port selection problem in fluid MIMO systems via low-complexity algorithms, and \cite{Shuaixin} designed a bit error rate (BER)-oriented optimization algorithm for point-to-point FA-MIMO systems. Subsequently, the research progressed to single-cell multi-user systems \cite{tang2025capacity,Tianyi,Chen2026,ISAC}. Specifically, \cite{tang2025capacity} studied an uplink multiuser system with FAs at both ends, and \cite{Tianyi} proposed a block coordinate ascent algorithm that jointly optimizes beamforming and the positions of both ends for downlink multi-user systems. FAS has also been integrated with other functionalities, such as low-latency communications \cite{Chen2026} and integrated communications and sensing (ISAC) \cite{ISAC}. Nevertheless, all of the aforementioned methods rely on numerical optimization, which is sensitive to initialization and prone to convergence to saddle points or local optima.

To combat this, researchers have applied deep reinforcement learning (DRL) to optimize FA positioning in single-cell multi-user systems \cite{Ho2025,zhang2026indoor,waqar2024opportunistic,weng2024learning,wang2024fluid}. Specifically, \cite{Ho2025} adopted proximal policy optimization (PPO) for latency minimization in the federated learning-assisted FAS, and 
\cite{zhang2026indoor} applied group relative policy optimization (GRPO) to indoor FAS optimization, reducing the model size and floating-point operations (FLOPs) by half compared with PPO. In addition, \cite{waqar2024opportunistic} developed a softmax deep deterministic
policy gradient (SD3) algorithm to jointly optimize opportunistic user scheduling and port selection. In the presence of imperfect channel state information (CSI), \cite{weng2024learning} applied a heterogeneous multi-agent deep deterministic policy gradient (MADDPG) algorithm in which separate agents are dedicated to MIMO beamforming and FA positioning. Besides, \cite{wang2024fluid} proposed an advantage actor-critic (A2C) to jointly optimize FA position and MIMO beamforming with ISAC.  To summarize, all of the aforementioned works establish DRL as an effective approach for tackling the non-convex, high-dimensional joint FA position and MIMO beamforming optimization problem.

Recent research has shifted to FA position optimization in FA-assisted wireless networks \cite{wei2025movable, zhu2025joint, Zhao2026FA,  mao2025joint, wu2026fluid, Ghadi2025, Li2026}. For cell-free networks, \cite{wei2025movable} proposed a dynamic neighborhood pruning particle swarm optimization (PSO) algorithm to minimize the maximum transmit power. In addition, \cite{zhu2025joint} developed a two-loop optimization framework combining PSO, fractional programming, and successive convex approximation for cell-free massive MIMO networks. \cite{Zhao2026FA} proposed an FA-assisted mobile edge computing (MEC) space-air-ground integrated network (SAGIN) system that jointly optimizes resource allocation and FA positioning to minimize task computation delay. \cite{mao2025joint} jointly optimized time scheduling and two-stage port activation for FA-assisted wireless powered communication networks. \cite{wu2026fluid} proposed a novel FA and stacked intelligent metasurface (SIM)-enabled multi-user ISAC system for low-altitude wireless networks. \cite{Ghadi2025} addressed the two-user FA multiple access (FAMA) in strong interference channels and adopted simultaneous non-unique decoding to turn strong interference into a useful resource. \cite{Li2026} achieved simultaneous enhancement of spectral and energy efficiency through alternating optimization of port selection, beamforming, and reconfigurable intelligent surface (RIS) phase shifts in FA-assisted cell-free networks.
Nevertheless, all of these works focus on numerical optimization methods.

By treating each BS as an agent, multi-agent reinforcement learning (MARL) has been used for FA-assisted wireless networks \cite{Li2025MATD3} and \cite{su2024cd}. In FA-assisted cell-free networks, \cite{Li2025MATD3} proposed a multi-agent twin delayed deep deterministic policy gradient (MATD3) algorithm within the centralized training with decentralized execution (CTDE) paradigm, based on each BS's local observations. In FA-assisted networks, \cite{su2024cd} introduced a multi-agent proximal policy optimization (MAPPO) algorithm for power allocation, deployed under the CTDE paradigm. All of these works show that, in FA-assisted networks, MARL achieves promising performance without requiring BS coordination. Nevertheless, the excessively long training time of MARL remains a significant challenge.

\subsection{Our Contributions}
In this paper, we propose a critic-free multi-agent group relative policy optimization (MAGRPO) algorithm. As summarized in Table~\ref{tab:related_work}, our work jointly optimizes FAS positions, beamforming, and transmit power using DRL. Our main contributions are summarized as follows:
\begin{itemize}
\item \textit{Dec-POMDP Problem Formulation}: We formulate the optimization of FA positions, beamforming, and power allocation in FA-assisted wireless networks as a decentralized partially observable Markov decision process (Dec-POMDP). This formulation captures the distributed decision-making architecture and partial observability of each BS during operation, and enables the application of MARL with the CTDE paradigm.

    \item \textit{MAGRPO Algorithm}: 
   For this problem, we propose MAGRPO, a critic-free CTDE algorithm that differs from MAPPO in both advantage estimation and policy updating. Specifically, MAGRPO substitutes MAPPO \cite{MAPPO}'s centralized critic and generalized advantage estimation (GAE) with a relative advantage derived from groups of joint trajectories, and employs this advantage to update actor policies. We further introduce a MAPPO warm-up stage for training stability and an entropy bonus for policy exploration. Under parameter sharing, MAGRPO reduces the computational complexity by approximately half.

    \item \textit{Experimental Validation}: We conduct extensive simulations in FA-assisted wireless networks to evaluate MAGRPO against MAPPO, MATD3, and a fixed-position baseline (MAPPO-FA). The results demonstrate that MAGRPO achieves sum-rate performance higher than MATD3 and comparable to or slightly higher than MAPPO while reducing the training time by about $20\%$--$23\%$ compared with MAPPO and by about $35\%$ compared with MATD3, and attains several-fold sum-rate gains over fixed-position configurations. Ablation studies further examine the effects of the group size, the trajectory length, and the MAPPO warm-up steps, providing practical guidance for hyperparameter selection.

\end{itemize}

\textit{Organization}: The remainder of this paper is organized as follows. Section II presents the system model and problem formulation. Section III recasts the problem as a Dec-POMDP and introduces the CTDE paradigm. Section IV reviews MAPPO, and Section V introduces the proposed MAGRPO. Section VI presents the simulation results, followed by concluding remarks in Section VII.

\textit{Notation}:   $(\cdot)^*$ denotes conjugate operation. $(\cdot)^H$ denotes conjugate transpose operation. $\mathbb{E}\{\cdot\}$ denotes the long-term expectation operator. We write $f(x) = \mathcal{O}(g(x))$ as $x \to \infty$ if there exist constants $C > 0$ and $x' \in \mathbb{R}$ such that $\lvert f(x) \rvert \le C \lvert g(x) \rvert$ for all $x \ge x'$. The notation $\|\cdot\|$ denotes the Euclidean norm for vectors and the spectral norm for matrices.
$\log$ denotes $\log_2$.  The clipping function is defined as $\text{clip}(x, 1-\epsilon, 1+\epsilon) := \max(\min(x, 1+\epsilon), 1-\epsilon)$.

\section{System Model and Problem Formulation}

\begin{figure}[t]
    \centering
    \includegraphics[width=0.95\linewidth]{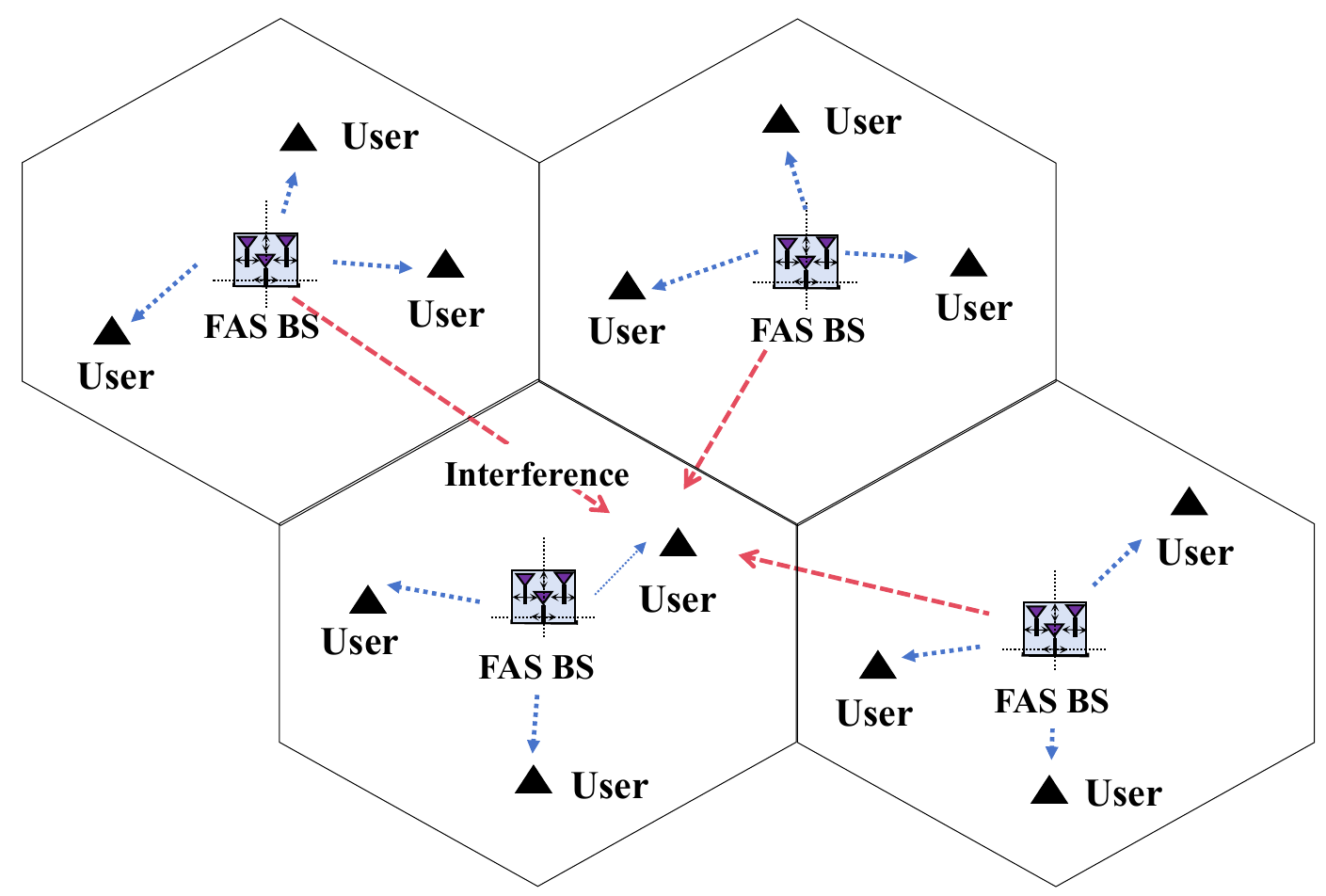}
    \caption{Illustration of an example of the considered downlink FA-assisted wireless network.}
    \label{fig:system_model}
\end{figure}

 We consider an FA-assisted wireless network in the downlink, where multiple BSs are equipped with two-dimensional (2D) FA arrays. In each cell, a BS $i \in \mathcal{N} \dis \{1,\cdots,N\}$ with $M$ fluid transmit antennas simultaneously serves $K$ single-antenna users. An example of the considered downlink FA-assisted wireless network is shown in Fig. \ref{fig:system_model}. The carrier wavelength and frequency are denoted by $\lambda$ and $f$, respectively. The transmit signal of BS $i$ is written as
\begin{equation}
\mathbf{x}_i = \sum_{k=1}^{K}\mathbf{w}^{[k]}_{i} s_{i}^{[k]},
\end{equation}
where $\mathbf{w}_{i}^{[k]}\in\mathbb{C}^{M\times 1}$ denotes the beamforming vector for user $k \in \mathcal{K} \dis \{1,\cdots,K\}$ from the BS $i$; and $s_{i}^{[k]}\in\mathbb{C}$ denotes the corresponding data symbol. We assume that all data streams are zero-mean, unit-power, and mutually uncorrelated, i.e.,
\begin{equation}
\mathbb E\{s_i^{[k]}\}=0,\qquad
\mathbb E\{s_i^{[k]}s_j^{[q]*}\}
=\begin{cases}
1, & (i,k)=(j,q),\\
0, & (i,k)\ne(j,q).
\end{cases}
\label{DataStreamAssumption}
\end{equation}
The input-output relationship at the user $k$ served by the BS $i$ is written as
\begin{align*}
  y_{i}^{[k]} = & \mathbf{h}_{i,i}^{[k]}\mathbf{w}_{i}^{[k]} s_{i}^{[k]} + \underbrace{\sum_{k'=1,k' \ne k}^{K}\mathbf{h}_{i,i}^{[k]}\mathbf{w}_{i}^{[k']} s_{i}^{[k']}}_{\text{intra-cell interference signal}}\nonumber \\
&  + \underbrace{\sum_{j=1,j \ne i}^{N}\sum_{k'=1}^{K}\mathbf{h}_{j,i}^{[k]}\mathbf{w}_{j}^{[k']} s_{j}^{[k']}}_{\text{inter-cell interference signal}} + z_{i}^{[k]}, \label{input-output}
\end{align*}
where $z_{i}^{[k]}\sim\mathcal{CN}(0,\sigma^2)$ denotes the additive white Gaussian noise (AWGN). Taking the expectation over the data symbols while conditioning on the channels and beamformers, the uncorrelated-stream assumption in \eqref{DataStreamAssumption} eliminates all cross terms. Hence, the average intra-cell and inter-cell interference powers are $I_{1,i}^{[k]} = \sum_{k'=1,k' \ne k}^{K}|\mathbf{h}_{i,i}^{[k]}\mathbf{w}_{i}^{[k']}|^2$ and $I_{2,i}^{[k]} = \sum_{j=1,j \ne i}^{N}\sum_{k'=1}^{K}|\mathbf{h}_{j,i}^{[k]}\mathbf{w}_{j}^{[k']}|^2$, respectively. The signal-to-interference-plus-noise ratio (SINR) is defined as the ratio of the desired-signal power to the total interference-plus-noise power. Accordingly, the SINR of user $k$ served by BS $i$ is
\begin{equation}
\gamma_{i}^{[k]}=
\frac{ \big| \mathbf{h}_{i,i}^{[k]}\mathbf{w}_{i}^{[k]} \big|^2 }
{I_{1,i}^{[k]} + I_{2,i}^{[k]}+\sigma^2}.
\label{eq:sinr_mc}
\end{equation}

\subsection{FAS Field Response Channel Model} 
As shown in the input-output relationship above, there are two kinds of channels. The channel to the intra-cell user, denoted by $\mathbf{h}_{i,i}^{[k]} \in \mathbb{C}^{1 \times M}$ for $i \in \mathcal{N}$ and $k \in \mathcal{K}$, is the channel from BS $i$ to its served user $k$. The channel to the inter-cell user, denoted by $\mathbf{h}_{i,j}^{[k]} \in \mathbb{C}^{1 \times M}$ for $i,j \in \mathcal{N}$ and $k \in \mathcal{K}$, is the channel from BS $i$ to user $k$ served by BS $j$ with $j \ne i$.  For brevity, we explain the model for a generic channel $\mathbf{h}$ and omit its indices. The specific indexed forms, such as $\mathbf{h}_{i,i}^{[k]}$ and $\mathbf{h}_{i,j}^{[k]}$, follow directly.

According to the field response model \cite{efrem2022transmit,Shuaixin,tang2025capacity,Tianyi}, the channel $\mathbf{h}$ is expressed as the product of the user's field response vector, the path-response matrix, and the BS's field response matrix. Specifically, the user's field response vector, denoted by $\mathbf{f}(\cdot)$, captures the impact of the angle of arrival (AoA) in modeling $\mathbf{h}$. We denote the user's antenna position by $\mathbf{v} \dis(x^r,y^r,z^r)$. Assume that there are $L$ paths from the BS to the user. The AoA for path $\ell \in \mathcal{L} \dis \{1,\cdots,L\}$ is denoted by $(\xi^{r}_{\ell}, \psi^{r}_{\ell})$. The virtual AoA components are then given by $\eta^{r}_{\ell}=\cos\xi^{r}_{\ell}\cos\psi^{r}_{\ell}$, $\beta^{r}_{\ell}=\cos\xi^{r}_{\ell}\sin\psi^{r}_{\ell}$, and $\nu^{r}_{\ell}=\sin\xi^{r}_{\ell}$. The user's field response vector $\mathbf{f}(\mathbf{v}) \in \mathbb{C}^{L \times 1}$ can be calculated by 
\begin{equation}
    \mathbf{f}(\mathbf{v}) = 
\big[e^{j\frac{2\pi}{\lambda}\rho_{1}(\mathbf{v})};\cdots;
e^{j\frac{2\pi}{\lambda}\rho_{L}(\mathbf{v})}\big],
\end{equation}
where $\rho_{\ell}(\mathbf{v}) = x^r\eta^{r}_{\ell}+y^r\beta^{r}_{\ell}+z^r\nu^{r}_{\ell}$.

The BS's field response matrix is composed of the BS's field response vectors. We denote the field response vector by $\mathbf{g}(\mathbf{u}_m),\, m \in \mathcal{M} \dis \{1,...,M\}$, where the two-dimensional position of the $m$-th FA at the BS is denoted by $\mathbf{u}_m \dis(x_m^t,y_m^t)$. The BS's field response matrix is expressed as $
\mathbf{G}(\mathbf{u}_{1},\cdots,\mathbf{u}_M) := \big[\mathbf{g}(\mathbf{u}_{1}),\cdots,\mathbf{g}(\mathbf{u}_{M})\big]
\in\mathbb{C}^{L \times M}.$ Specifically, the BS's field response vector
captures the impact of the angle of departure (AoD) in modeling $\mathbf{h}$. Since we assume that there are $L$ multipath components, the AoD for path $\ell \in \mathcal{L}$ and $m$-th FA position at BS is denoted by $(\xi^{t}_{\ell,m}, \psi^{t}_{\ell,m})$. The in-plane virtual AoD components are then given by $\eta^{t}_{\ell,m}=\cos\xi^{t}_{\ell,m}\cos\psi^{t}_{\ell,m}$ and $\beta^{t}_{\ell,m}=\cos\xi^{t}_{\ell,m}\sin\psi^{t}_{\ell,m}$. The field response vector at the $m$-th FA at BS $\mathbf{g}(\mathbf{u}_m) \in \mathbb{C}^{L \times 1}$ can be calculated by 
\begin{equation}
    \mathbf{g}(\mathbf{u}_m) =  \big[e^{j\frac{2\pi}{\lambda}\rho_{1}(\mathbf{u}_m)};\cdots;
e^{j\frac{2\pi}{\lambda}\rho_{L}(\mathbf{u}_m)}\big], 
\end{equation}
where $\rho_{\ell}(\mathbf{u}_m) = x_m^t\eta^{t}_{\ell,m}+y_m^t\beta^{t}_{\ell,m}$.

The path-response matrix represents the impact of multipath channel gain. Let $\mathbf{\Sigma} \in \mathbb{C}^{L \times L}$ be diagonal, with its $\ell$-th diagonal entry given by $\zeta_\ell/\sqrt{L}$ for $\ell \in \mathcal{L}$. We explicitly adopt the path-gain normalization $|\zeta_\ell|\le1$ for all $\ell\in\mathcal L$.
Consequently, $\|\boldsymbol\Sigma\|\le1/\sqrt L$. The channel vector $\mathbf{h}$ is then expressed as follows:
\begin{equation}
    \mathbf{h}
= 
\mathbf{f}^{H}(\mathbf{v})\,\mathbf{\Sigma}\,\mathbf{G}(\mathbf{u}_1,\cdots,\mathbf{u}_M).
\end{equation}

\subsection{Joint FA-Position and Beamforming Optimization}
\label{subsec:prob_mc}

For the considered downlink FA-assisted wireless network, we assume that there is no communication among BSs during decentralized execution. Let the random variable $\boldsymbol\xi$ denote the user locations. These locations are random samples rather than time-evolving states; hence, the system model does not assume user mobility or a physical user trajectory.

Let $\mathbf U\dis\{\mathbf u_m^i\}$ and $\mathbf W\dis\{\mathbf w_i^{[k]}\}$ denote the FA-position and beamforming optimization variables, respectively, where $i\in\mathcal N$, $m\in\mathcal M$, and $k\in\mathcal K$; the power allocated to user $k$ is embedded in $\|\mathbf w_i^{[k]}\|^2$. We first formulate the problem as a conventional stochastic optimization over random user locations:
\begin{subequations}
\begin{align}
\mathrm{P1}:\quad
\max_{\mathbf U,\mathbf W}\quad
&\mathbb E
\left[\sum_{i=1}^{N}\sum_{k=1}^{K}
\log\!\left(1+\gamma_i^{[k]}\right)\right]
\label{obj}\\
\mathrm{s.t.}\quad
&\sum_{k=1}^{K}\|\mathbf w_i^{[k]}\|^2\le P_{\max},
\label{P1-C1}\\
&\mathbf u_m^i\in\mathcal C,
\label{P1-C2}\\
&\|\mathbf u_m^i-\mathbf u_{m'}^i\|\ge D_{\min}.
\label{P1-C3}
\end{align}
\end{subequations}
Here, the expectation is taken over the distribution of $\boldsymbol\xi$. Constraint \eqref{P1-C1} applies to all $i\in\mathcal N$; \eqref{P1-C2} applies additionally to all $m\in\mathcal M$; and \eqref{P1-C3} applies to all distinct $m,m'\in\mathcal M$. Objective \eqref{obj} is the expected network sum-rate over the user-location distribution; \eqref{P1-C1} enforces the maximum transmit power; \eqref{P1-C2} defines the two-dimensional movable region $\mathcal C\subset\mathbb R^2$; and \eqref{P1-C3} imposes the minimum spacing between two FAs to prevent mutual coupling.

Problem P1 is challenging because it is a non-convex, high-dimensional stochastic optimization problem due to the field-response channel model and the minimum FA spacing constraint. We next reformulate P1 as a Dec-POMDP and solve it using MARL.

\section{Dec-POMDP and CTDE Paradigm}

%In this section, we first reformulate Problem P1 as a Dec-POMDP, and then introduce the CTDE paradigm for MARL.

\subsection{Dec-POMDP Reformation of Problem P1}
We use a Dec-POMDP to learn the distributed policy in Problem P1. This formulation captures the distribution of user locations, the iterative joint decisions used during MARL training, and the fact that each BS has only local observability and cannot communicate with other BSs during decentralized execution. The resulting MARL rollout is an algorithmic device for policy optimization and does not represent a physical trajectory of user locations. Specifically, the ingredients of the Dec-POMDP are defined as follows:

 \textit{\textbf{Agents}}: Let $\mathcal{N}$ denote the set of indices of all BSs, i.e., $\mathcal{N} \dis \{1, \cdots, N\}$. Each BS acts as an independent agent.

 \textit{\textbf{State}}: Let $\mathcal{S}$ denote the state space, which contains complete information of the wireless network. $\mathcal{S}$ is defined as follows:
    \begin{align}
 \!\!\!\!       \mathcal{S} \dis  & \big\{\mathbf{u}_{m}^i,  \mathbf{w}_{i}^{[k]}, \mathbf{v}_i^{[k]},  \mathbf{h}_{i,i}^{[k]},  \mathbf{h}_{j,i}^{[k]}, \log\!\big(1+\gamma_i^{[k]})   \nonumber \\
  \!\!\!\!       &  | \forall i \in \mathcal{N}, \forall k \in \mathcal{K}, \, \forall m \in \mathcal{M}, \, \forall j \ne i, j \in \mathcal{N}\big\}.
    \end{align} 
It can be seen that $\mathcal{S}$ includes all FA positions, beamforming vectors, users' antenna locations, channel vectors, and users' rates, which provides a global view.

 \textit{\textbf{Observation}}: Let $\mathcal{O}$ denote the union of all BSs' local observations, i.e., $\mathcal{O} = \prod_{i=1}^N \mathbf{o}^i$, where the local observation at BS $i$ is denoted by $\mathbf{o}^i$. We define $\mathbf{o}^i$ as follows: 
    \begin{align}
      \mathbf{o}^i  \dis   &  \big\{\mathbf{u}_{m}^i,   \mathbf{w}_{i}^{[k]},  \mathbf{v}_i^{[k]},    \mathbf{h}_{i,i}^{[k]},   I_{2,i}^{[k]}, \log (1+\gamma_i^{[k]}) \nonumber \\
     &  |\forall k \in \mathcal{K}, \, \forall m \in \mathcal{M}, \, \forall j \ne i, j \in \mathcal{N}\big\}.
    \end{align}
Compared with the state, each BS has only a local observation derived from its limited information. 
The local observation $\mathbf{o}^i$ captures the BS's internal configuration, including its FA positions and beamforming vectors, as well as its perceived environment, which consists of the served users' antenna locations, channel vectors, rates, and the inter-cell interference signal power. Note that the inter-cell interference signal power, also known as the interference temperature, is key to quantifying the influence from other BSs and thus enables coordination.

 \textit{\textbf{Action}}: Let $\mathcal{A}$ denote the union of all BSs' local action, i.e., $\mathcal{A} = \prod_{i=1}^N \mathbf{a}^i$, where the local action at BS $i$ is denoted by $\mathbf{a}^i$. We denote $\mathbf{a}^i$ as follows:
    \begin{equation}
        \mathbf{a}^i \dis \big\{\mathbf{u}_{m}^i, \mathbf{w}_{i}^{[k]} \mid \forall k \in \mathcal{K},\, \forall m \in \mathcal{M} \big\}.
    \end{equation}
    The local action contains the FA positions and beamforming vectors produced by policy $\pi^i$ at each step of Problem P1.

 \textit{\textbf{Reward}}: Let $R_t$ denote the network-wide reward at step $t$. For a feasible joint action, it is defined as
    \begin{equation}
        R_t = \sum_{i=1}^N \sum_{k=1}^{K}\log(1 + \gamma_{i,t}^{[k]}).
        \label{reward}
    \end{equation}
For every feasible action, $R_t$ is the sample counterpart of the sum-rate integrand in \eqref{obj}. Consequently, the sample-average reward used during MARL training provides a Monte Carlo objective for the expectation in P1. The index $t$ refers only to a step of the algorithmic rollout and does not index user motion. The reward is shared network-wide rather than assigned per BS. Constraint \eqref{P1-C1} is enforced by beamforming normalization, \eqref{P1-C2} by mapping the position action into $\mathcal C$, and \eqref{P1-C3} by resampling the FA positions until the minimum-spacing constraint is satisfied.

The above Dec-POMDP is consistent with Problem P1: its joint policy is the optimization variable, user locations follow the distribution of $\boldsymbol\xi$, and its sample-average reward estimates the expectation in \eqref{obj}.

\subsection{CTDE Paradigm for MARL}

To solve the Dec-POMDP formulation efficiently with MARL, we adopt the CTDE paradigm, which aligns perfectly with the Dec-POMDP and offers great advantages in relying solely on local observations.

The CTDE paradigm was introduced in \cite{CTDE}. It leverages centralized information during training to optimize policies that rely solely on local observations during execution. Specifically, the CTDE paradigm addresses non-stationarity through a centralized critic that provides stable gradients, resolves credit assignment by decomposing global feedback into individual updates, and overcomes communication constraints because agents cannot share signals during execution and must rely solely on their local observations. In what follows, we introduce MAPPO, a representative MARL algorithm that adopts the CTDE paradigm.

\section{MAPPO Algorithm}
\label{sec:mapppo}

We first briefly introduce MAPPO, which serves as the foundation for the proposed MAGRPO. According to \cite{MAPPO}, MAPPO employs the CTDE paradigm to condition policy updates on global state. Relying on a stochastic policy gradient, MAPPO can handle hybrid continuous and discrete action spaces. Meanwhile, by incorporating PPO's advantage clipping mechanism, MAPPO ensures stable training by preventing oscillations and policy collapse.

\subsection{MAPPO Training}
The MAPPO training procedure consists of three phases.

\textit{\textbf{Phase-I}: Trajectory Collection}.
All agents interact with the environment for $T$ steps according to their current policies, collecting trajectory data. During this process, each agent's actor makes decisions based solely on its local observations while the centralized critic can access the global state information.

\textit{\textbf{Phase-II}: GAE}. Based on the collected trajectories, a centralized critic network computes the advantage function $\widehat{A}_t$ via GAE. The equation is
\begin{equation}
    \widehat{A}_t = \sum_{l=0}^{\infty} (\gamma_d \lambda_s)^l \delta_{t+l}, \label{GAE1}
\end{equation}
where $\lambda_s \in [0,1]$ controls the bias-variance tradeoff in GAE.
The TD residual in \eqref{GAE1} is computed by
\begin{equation}
    \delta_{t}  = R_t + \gamma_d V_\phi (\mathbf{s}_{t+1}) - V_\phi (\mathbf{s}_t), \label{GAE2}
\end{equation}
where $R_t$ is the reward at step $t$, $V_\phi(\mathbf{s}_t)$ denotes the output of the critic network for the global state $\mathbf{s}_t$ with network parameters $\phi$; and $\gamma_d \ge 0$ is the discount factor. The resulting advantage functions $\{\widehat{A}_t\}_{t=1}^T$ used in actors should be normalized within each batch to have zero mean and unit variance.

\textit{\textbf{Phase-III}: Actors and Critic Networks Update}.
Based on GAE, MAPPO updates the centralized critic and the $N$ decentralized actor networks. An entropy bonus is also added to the policy objective to encourage sufficient exploration.

The critic is updated by minimizing a mean-squared error (MSE) loss, given by
\begin{equation}
\min_{\phi} \,\,\, \frac{1}{T} \sum_{t=1}^T \left(V_{\phi}(\mathbf{s}_t)-\hat{R}_t\right)^2, \label{critic_loss}
\end{equation}
where $\hat{R}_t$ is the bootstrapped reward estimate. This target is defined as $\hat{R}_t = \hat{A}_t + V_{\phi_{\text{old}}}(\mathbf{s}_t)$, where $\hat{A}_t$ is the advantage estimate and $V_{\phi_{\text{old}}}(\mathbf{s}_t)$ is the value produced by the network parameters from the previous step, ensuring the target is not dependent on the current parameters being updated.

 Given the local observation $\mathbf{o}^i_{t-1}$ at step $t-1$, we denote the action probability output by actor network $i$ as $\pi_{\theta^i}(\mathbf{o}_{t}^i\,|\,\textbf{a}^i,\mathbf{o}^i_{t-1})$ at step $t$, where $\theta^i$ denotes the network parameters. The $N$ actor networks are updated simultaneously by maximizing the sum of the PPO clipped surrogate objective and entropy bonus, given by
\begin{equation}
\max_{\{\theta^i\}_{i=1}^N}\  \sum_{i=1}^N (\mathcal{L}_{\pi}(\theta^i) + \iota\,\mathcal{L}_H(\theta^i)), \label{actorupdate}
\end{equation}
where $\mathcal{L}_{\pi}(\theta^i)$, given in \eqref{MAPPO-Obj}, represents the clipped policy gradient objective with advantage clipping threshold $\epsilon_1$ to prevent destructive policy updates, shown at the top of the next page; $\mathcal{L}_H(\theta^i)$ is the policy entropy, weighted by the entropy coefficient $\iota$ to encourage exploration. 
\begin{figure*}[!t]
    \begin{equation}
          \mathcal{L}_{\pi}(\theta^i)
\dis   \frac{1}{T}\sum_{t=1}^T  
\min\left\{
\frac{\pi_{\theta^i}(\mathbf{o}_t^i \,|\, \mathbf{a}^i, \mathbf{o}_{t-1}^i)}{\pi_{\theta^i_{\text{old}}}(\mathbf{o}_t^i \,|\, \mathbf{a}^i, \mathbf{o}_{t-1}^i)}\widehat{A}_t,\;
\text{clip}\!\left(\frac{\pi_{\theta^i}(\mathbf{o}_t^i \,|\, \mathbf{a}^i, \mathbf{o}_{t-1}^i)}{\pi_{\theta^i_{\text{old}}}(\mathbf{o}_t^i \,|\, \mathbf{a}^i, \mathbf{o}_{t-1}^i)},1-\epsilon_1,1+\epsilon_1\right)\widehat{A}_t
\right\}. \label{MAPPO-Obj}
    \end{equation}  
\end{figure*}
$\mathcal{L}_H(\theta^i)$ is given as follows:
\begin{equation}
\mathcal{L}_{H}(\theta^i)
\dis
\frac{1}{T}\sum_{t=1}^T
\mathcal{H}\!\left(\pi_{\theta^i}(\mathbf{o}_t^i\,|\,\mathbf{a}^i,  \mathbf{o}_{t-1}^i)\right),
\end{equation}
 where $\mathcal{H}(\cdot)$ denotes the differential entropy.

\begin{figure*}[!t]
\begin{equation*}
\resizebox{0.90\textwidth}{!}{$\displaystyle
    \!\!\!	\mathcal{J}_\pi(\theta^i) \dis  \frac{1}{G} \sum_{g=1}^G \frac{1}{T} \sum_{t=1}^{T} \left\{ 
		\min \left(
		\frac{\pi_{\theta^i}(\mathbf{o}_{t}^{i,g}|\textbf{a}^i,\mathbf{o}^{i,g}_{t-1})}{\pi_{\theta_{\text{old}}^i}(\mathbf{o}_t^{i,g}|\textbf{a}^i,\mathbf{o}^{i,g}_{t-1})}
		, \text{clip}\left(\frac{\pi_{\theta^i}(\mathbf{o}_t^{i,g}|\textbf{a}^i,\mathbf{o}^{i,g}_{t-1})}{\pi_{\theta^i_\text{old}}(\mathbf{o}_t^{i,g}|\textbf{a}^i,\mathbf{o}^{i,g}_{t-1})},1-\epsilon_2,1+\epsilon_2\right)\right)\widehat{A}_{g} - \mu \mathbb{D}_\text{KL}[\pi_{\theta^i}\|\pi_\text{ref}^i]
		\right\}$} \tag{21}\label{GRPO}
\end{equation*}
\hrule
\end{figure*}

\section{Proposed MAGRPO Algorithm}

Extending GRPO \cite{GRPO} from a single-agent to multi-agent is not a direct replication. Our proposed MAGRPO differs from GRPO \cite{GRPO} in four aspects. First, each group contains $G$ complete joint trajectories generated collectively by agents, and their network-wide cumulative returns produce a common network-level relative advantage. Second, this network-wide advantage weights each agent's locally clipped likelihood ratio, and the resulting actor objectives are summed to implement a joint policy update while preserving decentralized execution. Third, unlike GRPO \cite{GRPO}, we adopt MAPPO warm-up and KL regularization to stabilize simultaneous multi-agent updates after the centralized critic is removed. Fourth, beyond GRPO \cite{GRPO}, we incorporate an entropy bonus term in the loss function to facilitate exploration.

Since MAGRPO inherits the warm-up stage from MAPPO but dispenses with the centralized critic of MAPPO, we first contrast the two training architectures. For comparison, MAPPO updates the actors together with a centralized critic, which relies on GAE to estimate advantages and thus adds an extra network to every training step. MAGRPO removes the critic and GAE, and instead computes a group relative advantage from the returns of the collected joint trajectories to weight the updates of all actors.  The resulting training procedure is detailed in the following subsection.

\subsection{Proposed MAGRPO Training Algorithm}

%The proposed MAGRPO training process can be divided into four phases, described as follows:

The proposed MAGRPO training algorithm consists of the following sequential phases.

\textit{\textbf{Phase I}: MAPPO Warm-Up}.  
In this initial phase, we train MAPPO for a limited number of iterations to initialize the policies used in MAGRPO, producing a set of reference policies $\{\pi_\text{ref}^i\}_{i=1}^{N}$. These policies serve as a basis for guiding policy optimization in the subsequent training phases.

\textit{\textbf{Phase II}: Group Trajectory Collection}.  
All agents jointly collect a group of \(G\) rollouts of length \(T\) by interacting with the environment using their current policies. User locations are sampled from the distribution of $\boldsymbol\xi$ to form training samples. The rollout index organizes the sequence of policy updates and reward samples; it does not represent a physical user-location trajectory. The step-wise rewards are accumulated to obtain the rollout return. Due to CTDE, each agent's policy relies only on its local observations, thus enabling distributed execution.

\textit{\textbf{Phase-III}: Group Relative Advantage Estimate}. Based on the group of \(G\) trajectories, MAGRPO employs a group relative advantage estimate, thereby eliminating the need for a critic network. Specifically, the group relative advantage estimate for the $g$-th trajectory is calculated by
\begin{align}
 	 \widehat{A}_{g} =  \dfrac{\textstyle \sum_{t=1}^{T} R_t^g - \operatorname{Mean} \{\overbrace{\textstyle \sum_{t=1}^{T} R_t^1,\cdots,\textstyle \sum_{t=1}^{T} R_t^G}^{G\, \text{trajectories}}\}}{\operatorname{Std} \{\underbrace{\textstyle \sum _{t=1}^{T} R_t^1,\cdots,\textstyle \sum_{t=1}^{T} R_t^G}_{G\, \text{trajectories}} \}},   \label{GRPOAd} 
\end{align}
where \(R_t^g\) denotes the reward for the trajectory \(g\) at step \(t\); \(\operatorname{Mean}\{\cdot\}\) and \(\operatorname{Std}\{\cdot\}\) denote the mean and standard deviation, respectively. As shown in \eqref{GRPOAd}, the group relative advantages have zero sample mean and unit sample variance; this standardization does not require them to follow a Gaussian distribution. In contrast to the GAE used by MAPPO, the group relative advantage estimate of MAGRPO removes the need for a centralized critic network. 

\textit{\textbf{Phase-IV}: Actor Networks Update}. Using the group relative advantage estimate, MAGRPO updates the \(N\) actor networks by maximizing a weighted sum of the GRPO clipped surrogate objective and an entropy bonus.

\begin{algorithm}[t]
\caption{Proposed MAGRPO Training Algorithm}
\label{alg:magrpo}
\KwIn{$T_{\mathrm{ref}},\, T_{\max},\, T,\, G,\, \epsilon_1,\, \epsilon_2,\, \mu,\, \iota,\, E$}

Initialize replay buffer $\mathcal{D} \leftarrow \emptyset$, $T_\text{count} \leftarrow 0$\;

\textbf{Phase I: MAPPO Warm-Up}\;
\While{$T_\text{count} < T_{\text{ref}}$}{
    Collect trajectories by $\{\pi_{\theta^i}\}_{i=1}^N$, store in $\mathcal{D}$\;
    $T_\text{count} \leftarrow T_\text{count} + T$\;
    Compute $\widehat{A}_t$ via \eqref{GAE1}-\eqref{GAE2}\;
    \tcp{Estimate per-step advantages using GAE}
  \For{$e = 1$ \KwTo $E$}{
    Update the critic network by \eqref{critic_loss} via Adam\;
    Update $N$ actor networks by \eqref{actorupdate} via  Adam\;
}
    $\theta_{\text{old}}^i \leftarrow \theta^i$, $\forall i$\;
}
$\pi_{\text{ref}}^i \leftarrow \pi_{\theta^i}$, $\forall i$; $\mathcal{D} \leftarrow \emptyset$; $T_\text{count} \leftarrow 0$\;
\tcp{Save as reference policy}

\While{$T_\text{count} < T_{\max}$}{
    \textbf{Phase II: Group Trajectory Collection}\;
    \For{$g = 1$ \KwTo $G$}{
        Collect a trajectory by $\{\pi_{\theta^i}\}_{i=1}^N$, store in $\mathcal{D}$\;
    }
    $T_\text{count} \leftarrow T_\text{count} + G T$\;

    \textbf{Phase III: Group Relative Advantage Estimate}\;
    Compute $\widehat{A}_g$ via \eqref{GRPOAd}, $\forall g$\;

    \textbf{Phase IV: Actor Networks Update}\;
    \For{$e = 1$ \KwTo $E$}{
        Update $N$ actor networks by \eqref{MAGRPO} via Adam\;
       
    }
    $\theta_{\text{old}}^i \leftarrow \theta^i$, $\forall i$\;
}
\KwOut{Optimized policies $\{\pi_{\theta^i}\}_{i=1}^N$}
\end{algorithm}

Given the local observation $\mathbf{o}^i_{t-1}$ at step $t-1$, we denote the action probability output by actor network $i$ as $\pi_{\theta^i}(\mathbf{o}_{t}^{i,g}|\textbf{a}^i,\mathbf{o}^{i,g}_{t-1})$ for trajectory $g$ at step $t$, where $\theta^i$ denotes the network parameters. The
$N$ actor networks are updated simultaneously by maximizing
the sum of the GRPO clipped surrogate objective and entropy
bonus, given by
\begin{equation}
\max_{\{\theta^i\}_{i=1}^N}\  \sum_{i=1}^N (\mathcal{J}_\pi(\theta^i) + \iota \mathcal{J}_{H}(\theta^i)),  \label{MAGRPO} 
\end{equation}
where \(\mathcal{J}_\pi(\theta^i)\) denotes the GRPO objective for agent \(i\). Specifically, \(\mathcal{J}_\pi(\theta^i)\) is given in \eqref{GRPO}, 
shown at the top of the next page, which incorporates a KL divergence term and the advantage clipping threshold $\epsilon_2$. The strength of this regularization is controlled by the KL divergence penalty coefficient \(\mu\). According to \cite{GRPO}, KL divergence can be approximated by
\begin{align*}
	    \mathbb{D}_{\text{KL}}[\pi_{\theta^i}\|\pi_{\text{ref}}^i] \approx &\frac{\pi_{\text{ref}}^i(\mathbf{o}_{t}^{i,g}|\mathbf{a}^i,\mathbf{o}^{i,g}_{t-1})}{\pi_{\theta^i}(\mathbf{o}_t^{i,g}|\mathbf{a}^i,\mathbf{o}^{i,g}_{t-1})}  -  \nonumber \\
      & \ln \frac{\pi_{\text{ref}}^i(\mathbf{o}_t^{i,g}|\mathbf{a}^i,\mathbf{o}^{i,g}_{t-1})}{\pi_{\theta^i}(\mathbf{o}_t^{i,g}|\mathbf{a}^i,\mathbf{o}^{i,g}_{t-1})} - 1. \tag{22}
\end{align*}
\setcounter{equation}{22}
In addition, the entropy bonus $\mathcal{L}_{H}(\theta^i)$ is given as follows:
\begin{equation}
\mathcal{J}_{H}(\theta^i)
\dis
\frac{1}{G} \sum_{g=1}^G \frac{1}{T} \sum_{t=1}^{T} 
\mathcal{H}\!\left(\pi_{\theta^i}(\mathbf{o}_t^i\,|\,\mathbf{a}^i,  \mathbf{o}_{t-1}^i)\right).
\end{equation}

Overall, the proposed MAGRPO training algorithm is summarized in Algorithm 1. Adam optimizer \cite{kingma2014adam} can be applied to optimize the gradient descent.

 \subsection{Parameter Sharing and Computational Complexity}
\label{subsec:complexity}

Under parameter sharing, all agents reuse the same actor network, which maps each local observation to an action, with agent identity optionally encoded by an agent-specific one-hot vector. This design improves sample and training efficiency, supports a variable number of agents, simplifies training, and reduces computational and memory costs \cite{PS}.
 
In addition, we model both the $N$ actor networks and the single critic network as multilayer perceptrons with $J_{\text{hidden}}$ hidden layers of $J$ neurons each. Let $d_o$, $d_a$, and $d_s$ denote local observation, local action, and state dimensions, respectively. The computational complexity of one forward pass through the actor and critic networks is $
\mathcal{O}\big(J d_o + J d_a + J^2 J_{\text{hidden}}\big)
$ and $
\mathcal{O}\big(J d_s + J + J^2 J_{\text{hidden}}\big)
$, respectively. 

Under parameter sharing, the computational complexity of MAPPO per training step can be expressed as $\mathcal{O}(J(d_s + d_o + d_a) + J + 2J^2J_\text{hidden})$, which equals the sum of the forward-pass complexities of the actor and critic networks. The computational complexity of the proposed MAGRPO per training step can be expressed as $
\mathcal{O}\big(J d_o + J d_a + J^2 J_{\text{hidden}} + G\big)
$. Based on the state, local observation, and local action definitions in Section III-A, we have
$d_s = N\big[2M + 2KM(N+1) + 4K\big],  
d_o  = 2M + 4KM + 5K,  
d_a  = 2M + 2KM$. 
Given that $G \ll J d_s + J + J^2 J_{\text{hidden}}$, the proposed MAGRPO reduces the per-step computational complexity by approximately half compared with MAPPO.

\begin{table}[t]
	\centering
	\caption{Configuration and Hyperparameters}
	\label{Tab2}
	\footnotesize
	\setlength{\heavyrulewidth}{1pt}
	\setlength{\tabcolsep}{2pt}
	\renewcommand{\arraystretch}{1.08}
	\begin{tabular*}{0.96\columnwidth}{@{\extracolsep{\fill}}lccc@{}}
		\toprule
		\multicolumn{4}{c}{\textbf{Actor Network (MAGRPO, MAPPO)}} \\
		\textbf{Component} & \textbf{Type} & \textbf{Input/Output} & \textbf{Activation}\\
		Input Layer       & Linear      & Input Size$\times$256    & ReLU\\ 
		Hidden Layer 1     & Linear      & 256$\times$256           & ReLU \\  
        Hidden Layer 2    & Linear      & 256$\times$256           & ReLU \\  
		Output Layer      & Linear      & 256$\times$Action Size  & --     \\ 
		\addlinespace[2pt]
		\midrule
        \multicolumn{4}{c}{\textbf{Critic Network (MAPPO)}} \\
		\textbf{Component} & \textbf{Type} & \textbf{Input/Output} & \textbf{Activation}\\
		Input Layer       & Linear      & Input Size$\times$256    & ReLU\\ 
		Hidden Layer 1     & Linear      & 256$\times$256           & ReLU \\  
        Hidden Layer 2     & Linear      & 256$\times$256           & ReLU \\  
		Output Layer      & Linear      & 256$\times$1  & --     \\ 
		\addlinespace[2pt]
		\midrule
        \multicolumn{4}{c}{\textbf{Hyperparameters (MAGRPO, MAPPO)}} \\
        \textbf{Learning Rate}
        & \textbf{KL Penalty}
        & \textbf{Clipping Threshold}
        & \textbf{Batch Size} \\
        ``adaptive'' & $1 \times 10^{-6}$ & $0.1$ & $16$ \\
		\bottomrule
	\end{tabular*}%
\end{table}

\begin{figure*}[t]
    \centering
    \begin{subfigure}[b]{0.245\textwidth}
        \centering
        \includegraphics[width=\linewidth]{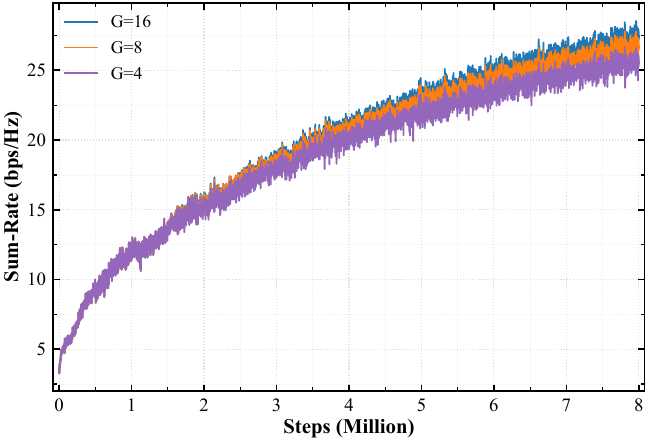}
        \caption{Training, $M=5$}
        \label{fig:group_m5}
    \end{subfigure}
    \hfill
    \begin{subfigure}[b]{0.245\textwidth}
        \centering
        \includegraphics[width=\linewidth]{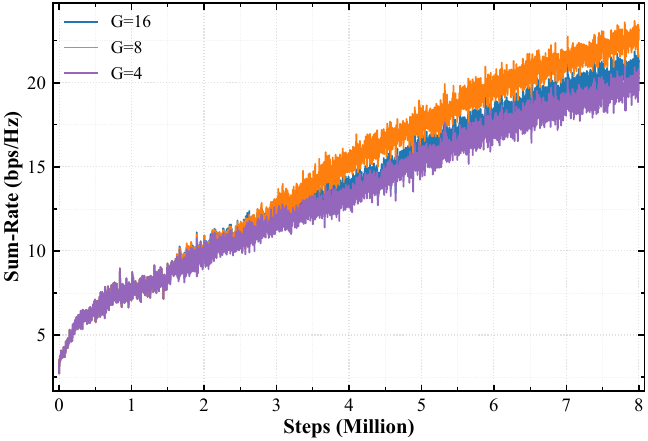}
        \caption{Training, $M=4$}
        \label{fig:group_m4}
    \end{subfigure}
    \hfill
    \begin{subfigure}[b]{0.245\textwidth}
        \centering
        \includegraphics[width=\linewidth]{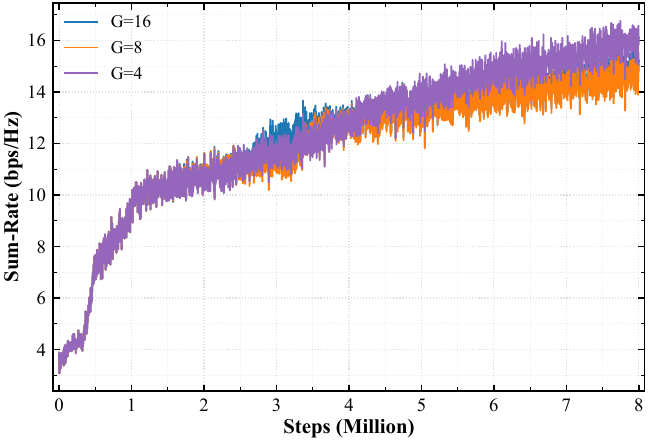}
        \caption{Training, $M=3$}
        \label{fig:group_m3}
    \end{subfigure}
    \hfill
    \begin{subfigure}[b]{0.245\textwidth}
        \centering
        \includegraphics[width=\linewidth]{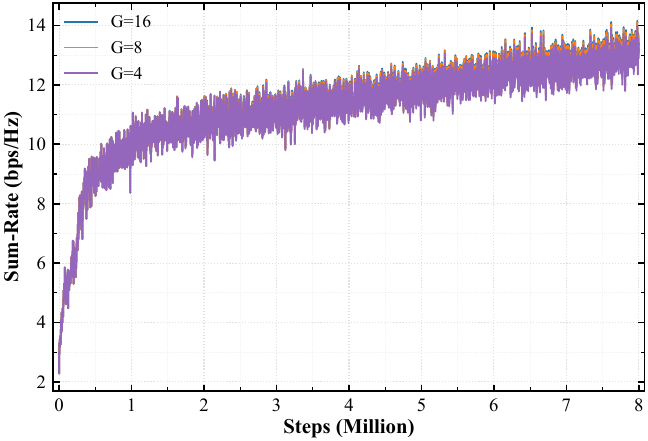}
        \caption{Training, $M=2$}
        \label{fig:group_m2}
    \end{subfigure}

    \begin{subfigure}[b]{0.245\textwidth}
        \centering
        \includegraphics[width=\linewidth]{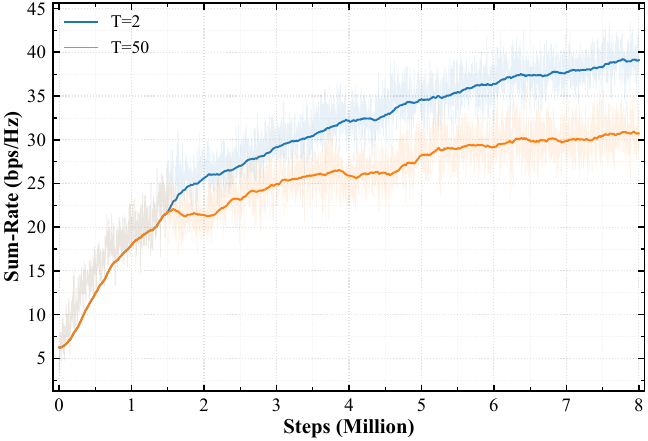}
        \caption{Testing, $M=5$}
        \label{fig:traject_m5}
    \end{subfigure}
    \hfill
    \begin{subfigure}[b]{0.245\textwidth}
        \centering
        \includegraphics[width=\linewidth]{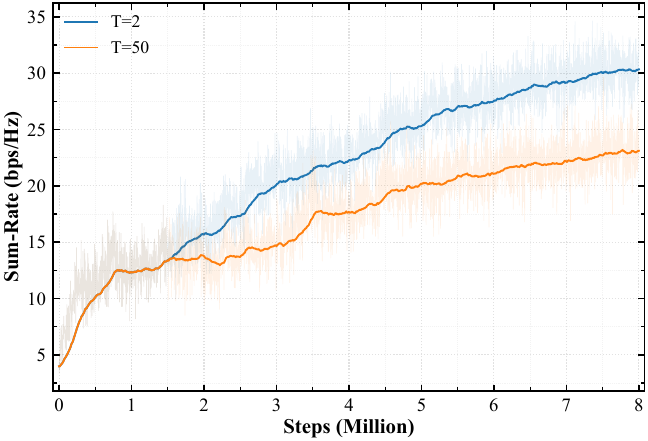}
        \caption{Testing, $M=4$}
        \label{fig:traject_m4}
    \end{subfigure}
    \hfill
    \begin{subfigure}[b]{0.245\textwidth}
        \centering
        \includegraphics[width=\linewidth]{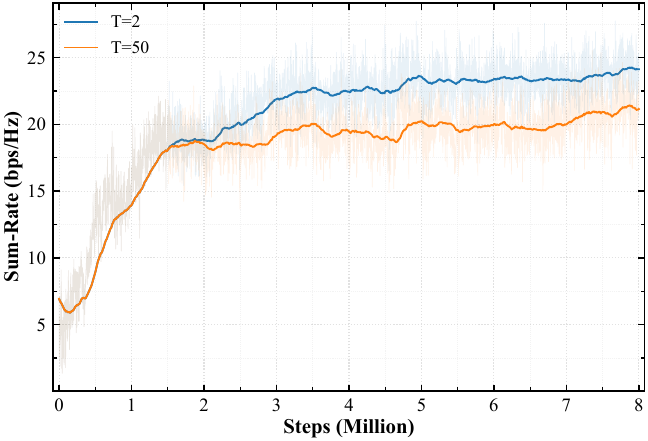}
        \caption{Testing, $M=3$}
        \label{fig:traject_m3}
    \end{subfigure}
    \hfill
    \begin{subfigure}[b]{0.245\textwidth}
        \centering
        \includegraphics[width=\linewidth]{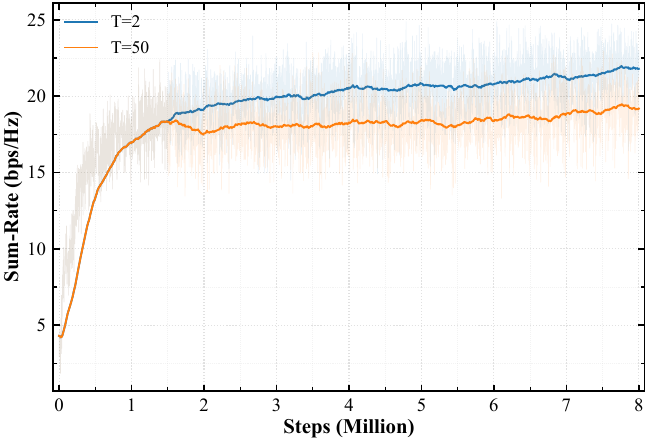}
        \caption{Testing, $M=2$}
        \label{fig:traject_m2}
    \end{subfigure}

    \caption{Influence of group size and trajectory length on MAGRPO.}
    \label{FGT}
\end{figure*}

\section{Simulations}

%In this section, we first present the simulation settings and baselines, and then the simulation results and discussion. 

\subsection{Simulation Settings and Baselines}
Unless specified otherwise, the simulation settings are as follows: the AWGN variance
is $\sigma^2 = - 91$~dBm. The maximum transmit power is $1$~W. The movable region for FAs is $0.5 \times 0.5$~$\text{m}^2$. The frequency and wavelength are $5.5$~GHz and $0.0545$~m, respectively. The minimum spacing between two FAs is $0.02$~m.
We employ parameter sharing as detailed in Section~V-B. 
When deploying the trained actor networks, the exact locations of all users across BSs cannot be predetermined, as the possible combinations of user locations are enormous. Accordingly, consistent with the user-location distribution in Problem P1, user positions are sampled from prescribed spatial distributions during both training and testing. These samples enable the policies to learn over a distribution of user locations rather than a single deterministic instance. The sampling does not model user movement. Specifically, user 1 is sampled from a sector spanning $3.5\sim5.5$~m in range and $80^\circ\sim90^\circ$ in azimuth, while user 2 is sampled from a sector spanning $3.5\sim5.5$~m in range and $90^\circ\sim100^\circ$ in azimuth. The BS is located $10$~m above the ground, and each user is located $1.8$~m above the ground. The horizontal distance between two BSs is $35$~m. 
All experiments were conducted on an NVIDIA RTX PRO 6000 Blackwell Workstation Edition GPU with $96$ GB of memory, an AMD Ryzen 9 9950X 16-Core Processor with $32$ logical processors, and $64$ GB of system memory. We use linear entropy annealing with the entropy coefficient decaying linearly from $0.01$ to $0.0015$ over $8$ million steps. Table \ref{Tab2} summarizes the neural network configurations for MAPPO and the proposed MAGRPO. Here, ``adaptive'' represents a dynamic learning rate schedule, i.e., starting at $3 \times 10^{-5}$ and held constant for the first 1400 updates, after which the learning rate is multiplied by a decay factor whenever the loss plateaus, with a minimum learning rate of $6 \times 10^{-6}$.  

The baseline algorithms used for comparison with the proposed MAGRPO are listed as follows.

\begin{figure*}[t]
    \centering
    % Row 1
    \begin{subfigure}[b]{0.245\textwidth}
        \centering
        \includegraphics[width=\linewidth]{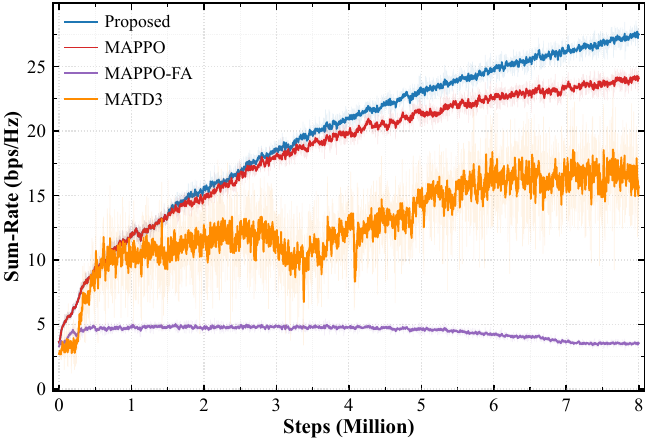}
        \caption{Training, $M=5$}
        \label{fig:error_ant_m5}
    \end{subfigure}
    \hfill    
    \begin{subfigure}[b]{0.245\textwidth}
        \centering
        \includegraphics[width=\linewidth]{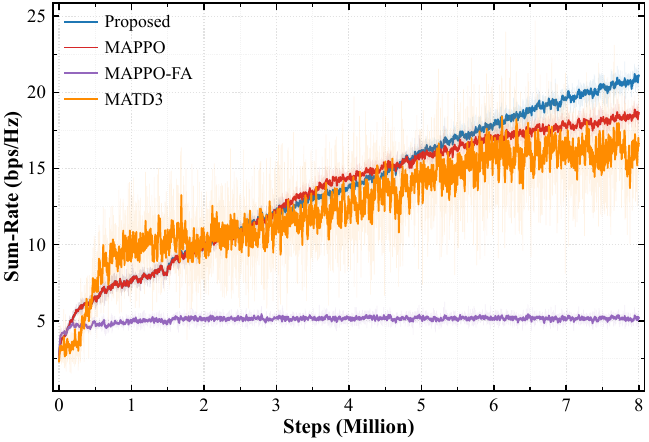}
        \caption{Training, $M=4$}
        \label{fig:error_ant_m4}
    \end{subfigure} 
     \hfill
     \begin{subfigure}[b]{0.245\textwidth}
        \centering
        \includegraphics[width=\linewidth]{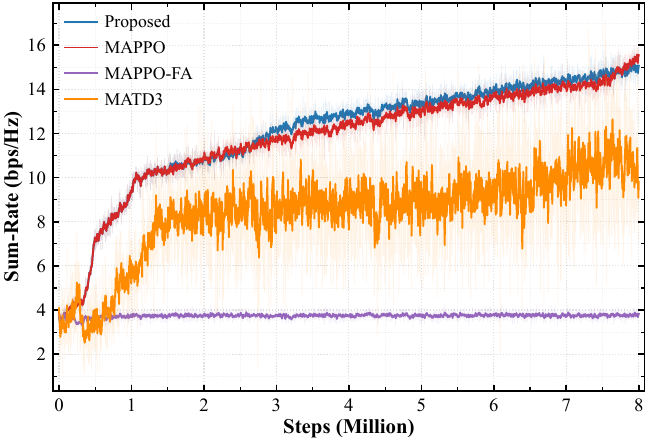}
        \caption{Training, $M=3$}
        \label{fig:error_ant_m3}
    \end{subfigure}
  \hfill
  \begin{subfigure}[b]{0.245\textwidth}
        \centering
        \includegraphics[width=\linewidth]{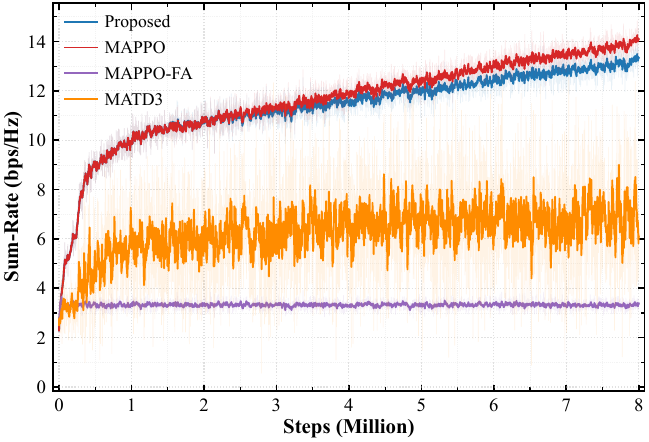}
        \caption{Training, $M=2$}
        \label{fig:error_ant_m2}
    \end{subfigure}

    % Row 2
   
     \begin{subfigure}[b]{0.245\textwidth}
        \centering
        \includegraphics[width=\linewidth]{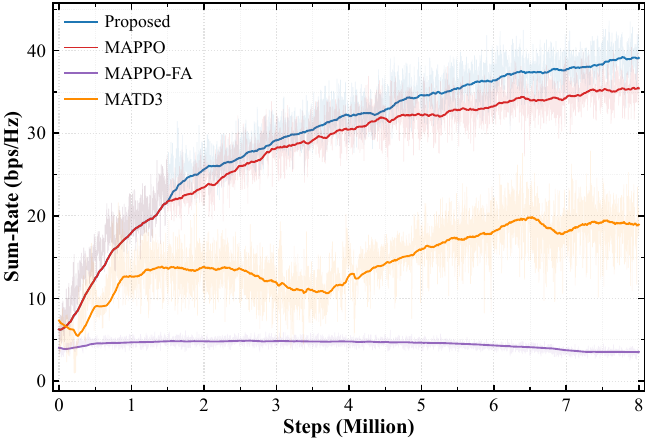}
        \caption{Testing, $M=5$}
        \label{fig:time_ant_m5}
    \end{subfigure}
    \hfill  
    \begin{subfigure}[b]{0.245\textwidth}
        \centering
        \includegraphics[width=\linewidth]{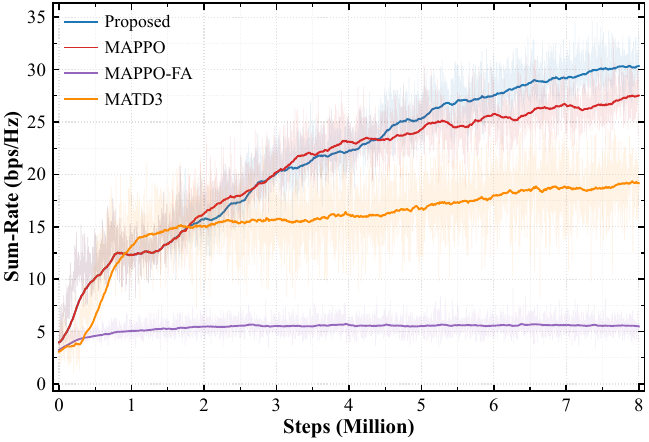}
        \caption{Testing, $M=4$}
        \label{fig:time_ant_m4}
    \end{subfigure}
         \hfill
    \begin{subfigure}[b]{0.245\textwidth}
        \centering
        \includegraphics[width=\linewidth]{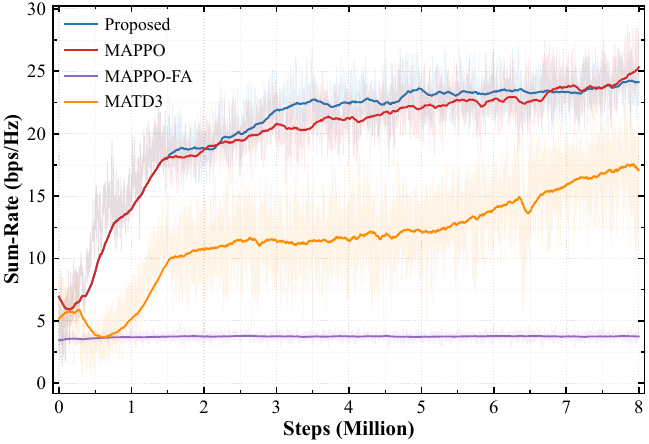}
        \caption{Testing, $M=3$}
        \label{fig:time_ant_m3}
    \end{subfigure}
    \hfill    
      \begin{subfigure}[b]{0.245\textwidth}
        \centering
        \includegraphics[width=\linewidth]{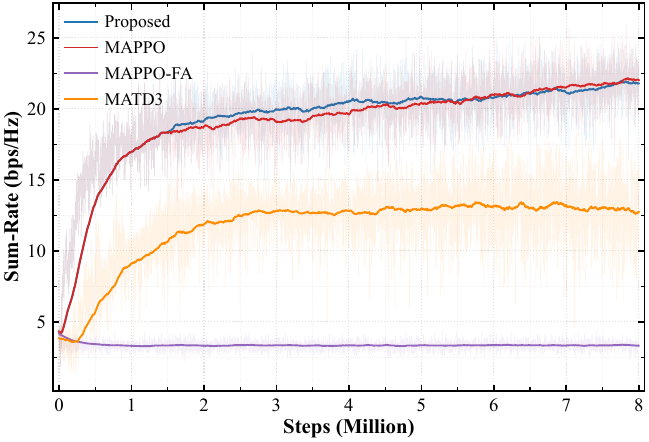}
        \caption{Testing, $M=2$}
        \label{fig:time_ant_m2}
    \end{subfigure}

    \begin{subfigure}[t]{0.245\textwidth}
        \vspace{0pt}
        \centering
        \includegraphics[width=\linewidth]{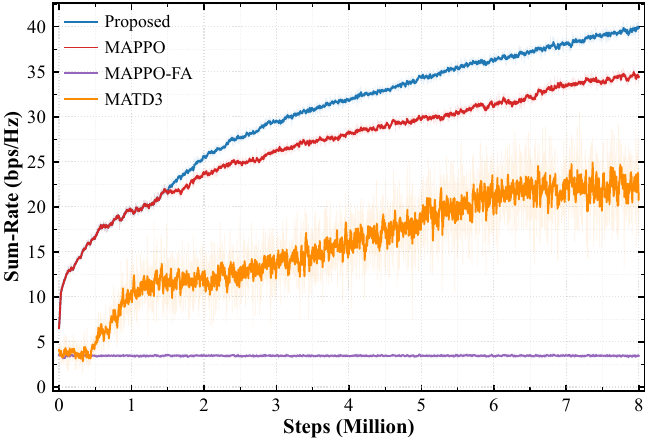}
        \caption{Training, $N=4$}
        \label{fig:train_n4}
    \end{subfigure}\hfill
    \begin{subfigure}[t]{0.245\textwidth}
        \vspace{0pt}
        \centering
        \includegraphics[width=\linewidth]{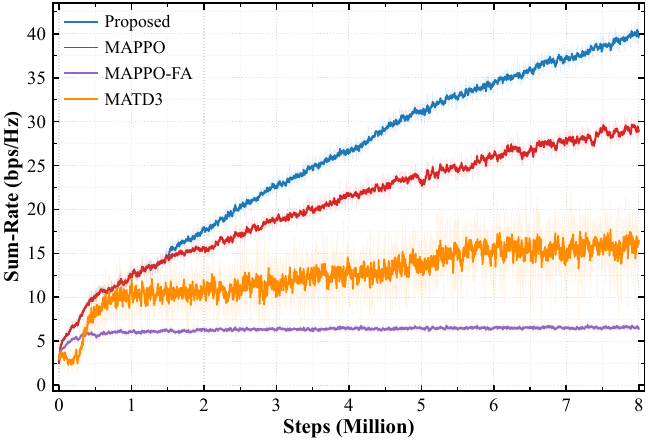}
        \caption{Training, $N=3$}
        \label{fig:train_n3}
    \end{subfigure}\hfill
    \begin{subfigure}[t]{0.245\textwidth}
        \vspace{0pt}
        \centering
        \includegraphics[width=\linewidth]{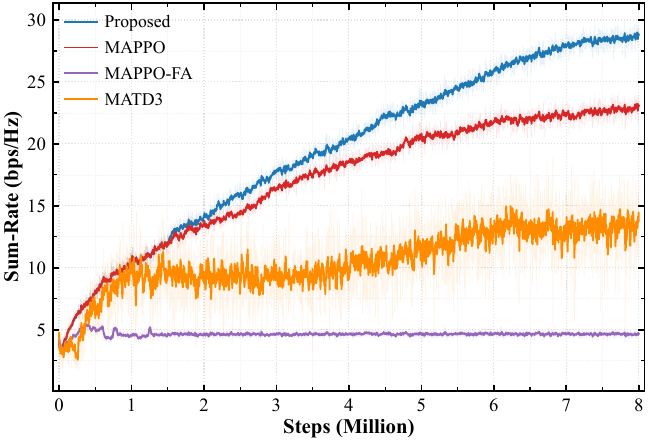}
        \caption{Training, $N=2$}
        \label{fig:train_n2}
    \end{subfigure}\hfill
    \begin{subfigure}[t]{0.245\textwidth}
        \vspace{0pt}
        \centering
        \includegraphics[width=\linewidth]{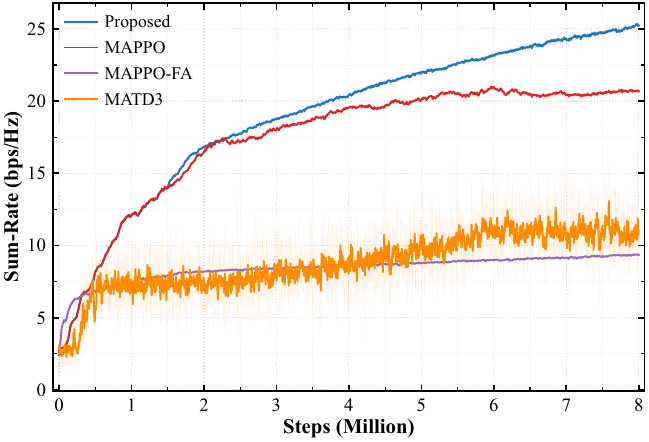}
        \caption{Training, $N=1$}
        \label{fig:train_n1}
    \end{subfigure}

     \begin{subfigure}[t]{0.245\textwidth}
        \vspace{0pt}
        \centering
        \includegraphics[width=\linewidth,height=0.72\linewidth]{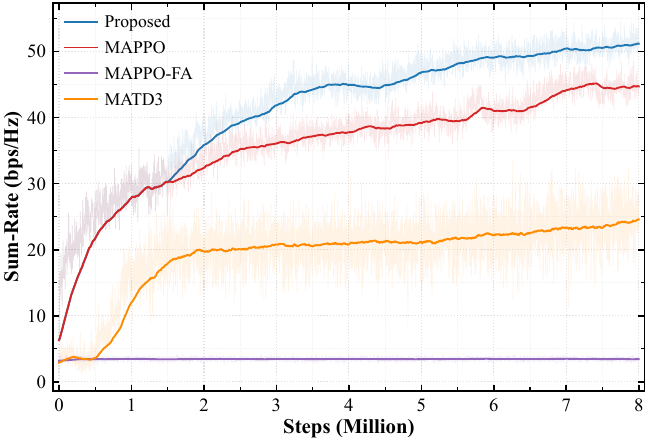}
        \caption{Testing, $N=4$}
        \label{fig:test_n4}
    \end{subfigure}\hfill
    \begin{subfigure}[t]{0.245\textwidth}
        \vspace{0pt}
        \centering
        \includegraphics[width=\linewidth,height=0.72\linewidth]{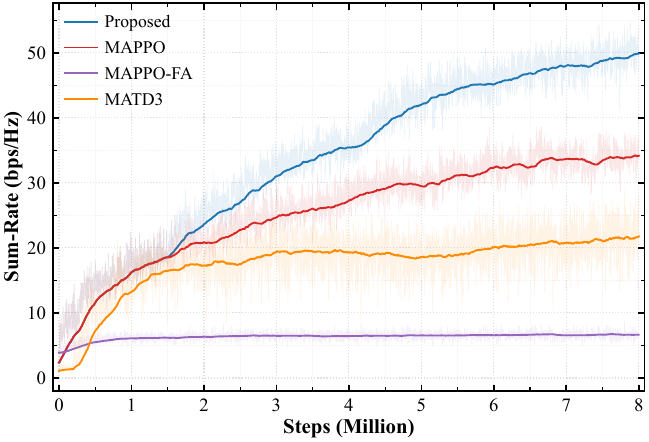}
        \caption{Testing, $N=3$}
        \label{fig:test_n3}
    \end{subfigure}\hfill
    \begin{subfigure}[t]{0.245\textwidth}
        \vspace{0pt}
        \centering
        \includegraphics[width=\linewidth,height=0.72\linewidth]{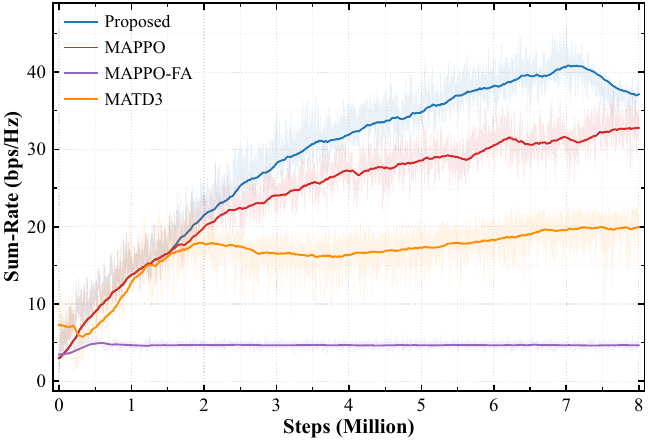}
        \caption{Testing, $N=2$}
        \label{fig:test_n2}
    \end{subfigure}\hfill
    \begin{subfigure}[t]{0.245\textwidth}
        \vspace{0pt}
        \centering
        \includegraphics[width=\linewidth,height=0.72\linewidth]{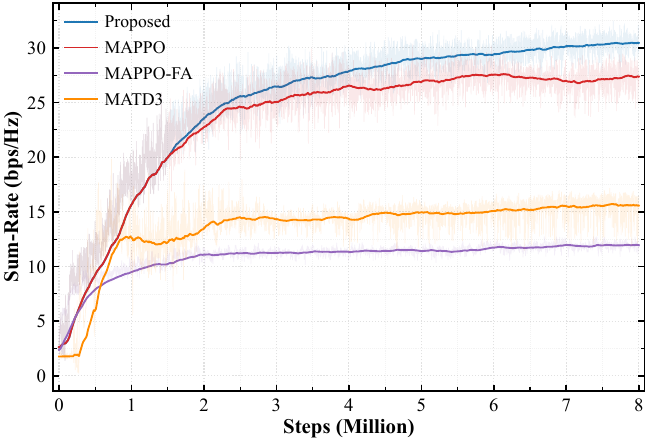}
        \caption{Testing, $N=1$}
        \label{fig:test_n1}
    \end{subfigure}
    \caption{Training and testing curves for the proposed MAGRPO and the baselines over varying $M$ and $N$.}
    \label{fig:compare_mn}
\end{figure*}

\begin{figure*}
    \centering 
    \begin{subfigure}[t]{0.245\textwidth}
        \vspace{0pt}
        \centering
        \includegraphics[width=\linewidth]{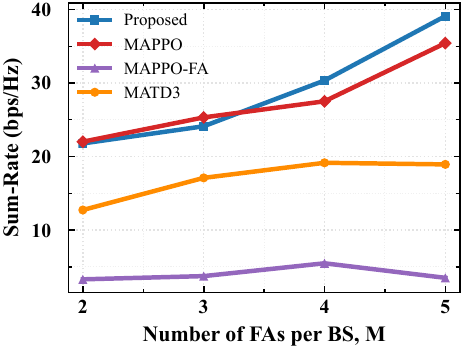}
        \caption{Sum-rate comparison, $M$}
        \label{fig:performace_m}
    \end{subfigure}\hfill
    \begin{subfigure}[t]{0.245\textwidth}
        \vspace{0pt}
        \centering
        \includegraphics[width=\linewidth]{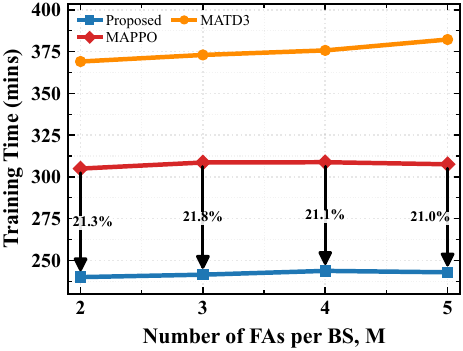}
        \caption{Training time comparison, $M$}
        \label{fig:time_m}
    \end{subfigure}\hfill
    \begin{subfigure}[t]{0.245\textwidth}
        \vspace{0pt}
        \centering
        \includegraphics[width=\linewidth]{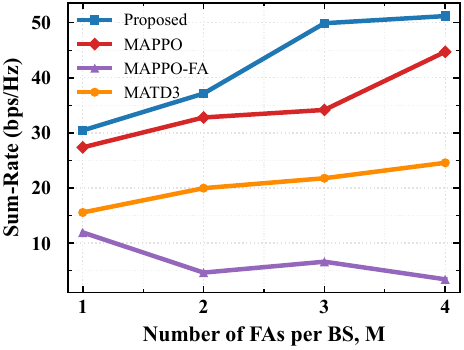}
        \caption{Sum-rate comparison, $N$}
        \label{fig:performance_n}
    \end{subfigure}\hfill
    \begin{subfigure}[t]{0.245\textwidth}
        \vspace{0pt}
        \centering
        \includegraphics[width=\linewidth]{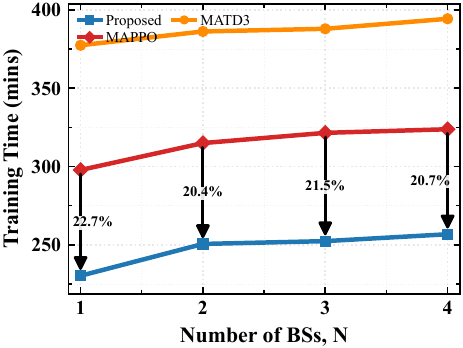}
        \caption{Training time comparison, $N$}
        \label{fig:time_n}
    \end{subfigure}

    \caption{Proposed MAGRPO over baselines over varying $M$ and $N$.}
    \label{fig:performace_mn}
\end{figure*}

\textbf{MAPPO}: This baseline implements MAPPO as 
proposed in \cite{MAPPO} and trains it for $8$ million steps.  Note that MAPPO, introduced in Section IV, serves as the foundation of the proposed MAGRPO. This baseline evaluates the sum rate, running time, and memory usage relative to the proposed MAGRPO.

\textbf{MATD3}: This baseline implements the multi-agent twin delayed deep deterministic policy gradient (MATD3) proposed in \cite{Ackermann2019}, which employs twin centralized critics, delayed policy updates, and target-policy smoothing to reduce overestimation and stabilize training. This algorithm has also been applied to FA-assisted cell-free networks \cite{Li2025MATD3}. It is trained for $8$ million steps, providing a comparison with a more robust deterministic CTDE method.

\textbf{MAPPO-FA}: This baseline freezes the FA positions to random values and optimizes only the beamforming by training MAPPO for $8$ million steps. In \cite{zhang2025multi}, MAPPO-FA has been used to optimize multi-cell spectrum and power allocation. This baseline evaluates the sum-rate improvement achieved through FA position optimization relative to MAPPO.

\subsection{Simulation Results and Discussion}

\begin{table}[t]
\centering
\caption{Impact of MAPPO Warm-Up Steps on MAGRPO.}
\label{tab:warmup_ablation}
%\footnotesize
\setlength{\tabcolsep}{4pt}
\renewcommand{\arraystretch}{1.08}
\begin{tabular}{ccc}
\toprule
\makecell{\textbf{MAPPO}\\\textbf{Warm-Up Steps}}
& \makecell{\textbf{MAGRPO Testing}\\\textbf{Sum-Rate (bps/Hz)}}
& \makecell{\textbf{Training Time Reduction}\\\textbf{over MAPPO}} \\
\midrule
0.5 million & 17.15 & $27.9\%$ \\
1 million   & 23.60 & $25.3\%$ \\
1.5 million & 30.35 & $21.1\%$ \\
2 million   & 31.37 & $20.1\%$ \\
\bottomrule
\end{tabular}
\end{table}

Table~\ref{tab:warmup_ablation} evaluates the effect of the MAPPO warm-up duration on the testing performance and training efficiency of MAGRPO. The training-time reduction is computed relative to training MAPPO for $8$ million steps and includes the MAPPO warm-up time. Increasing the number of warm-up steps provides a better-initialized policy and improves the training stability of the subsequent MAGRPO stage, but it also reduces the overall training-time saving. Thus, this ablation quantifies the tradeoff between testing sum-rate and end-to-end training efficiency. Thereby, we choose $1.5$ million warm-up steps for our proposed MAGRPO in the follow-up simulations.

In Fig. \ref{FGT}, we evaluate the influence of the group size and the trajectory length on the proposed MAGRPO. Fig. \ref{FGT} shows that, when the group size is relatively small, the training curves of MAGRPO exhibit significant fluctuations in sum rate.
 We attribute these fluctuations to the inaccurate estimate of the mean trajectory return used to center the relative-advantage numerator: with only a few trajectories, a single high- or low-return rollout can strongly bias the group mean and, consequently, the sign of the advantage. As the group size grows, the sample mean becomes more reliable, the advantage estimates become more stable, and the training curves become noticeably smoother. However, enlarging the group also consumes more environment steps before each policy update, so the smoothing benefit gradually saturates once the group is large enough to estimate the baseline reliably. This behavior is consistent across all evaluated antenna settings $M=2,3,4,5$, indicating that a moderate group size is sufficient in practice. Nevertheless, we choose $G=16$ in the follow-up simulations.

\begin{table*}[t]
    \centering
    \caption{Reduction Comparison of MAGRPO over Baselines}
    \label{tab:runtime_reduction}
    %\footnotesize
    \renewcommand{\arraystretch}{1.08}
    \setlength{\tabcolsep}{2pt}
    \begin{tabular*}{\textwidth}{@{\extracolsep{\fill}}cccccccccc@{}}
        \toprule
        \multirow{2}{*}{\raisebox{-0.8ex}{\textbf{System Setting}}}
        & \multicolumn{2}{c}{\textbf{Training Time Reduction}}
        & \multicolumn{2}{c}{\textbf{Model Size Reduction}}
        & \multicolumn{2}{c}{\textbf{FLOPs Reduction}} \\
        \cmidrule(lr){2-7}
        & \textbf{MAPPO} & \textbf{MATD3}
        & \textbf{MAPPO} & \textbf{MATD3}
        & \textbf{MAPPO} & \textbf{MATD3} \\
        \midrule
        $M=2,\,N=2$ & 21.3\% & 35\% & 52.6\% & 70.4\% & 52.4\% &  70.4\% \\
        $M=3,\,N=2$ & 21.8\% & 35.3\% & 53.3\% & 71.5\% & 53.4\% &  71.5\% \\
        $M=4,\,N=2$ & 21.1\% & 35.1\% & 54.0\% & 72.5\% & 54.0\% & 72.5\% \\
        $M=5,\,N=2$ & 21.0\% & 36.5\% & 54.6\% & 73.3\% & 54.6\% & 73.3\% \\
        $M=6,\,N=1$ & 22.7\% & 38.5\% & 47.4\% & 66.6\% & 47.4\% & 66.6\%  \\
        $M=6,\,N=2$ & 20.4\% & 35.0\% & 55.1\% & 74.0\% & 55.1\% & 74.0\%  \\
        $M=6,\,N=3$ & 21.5\% & 34.9\% & 62.7\% & 79.8\% & 62.8\% & 79.8\%  \\
        $M=6,\,N=4$ & 20.7\% & 34.9\% & 69.4\% & 84.1\% & 69.4\% & 84.2\% \\
        \bottomrule
    \end{tabular*}
\end{table*}

Moreover, Fig. \ref{FGT} shows that, as the trajectory length $T$ increases, the testing sum rate of MAGRPO decreases for all values of $M$. Here, $T$ is an algorithmic hyperparameter used to construct the group-relative advantage and does not model user mobility. We attribute this degradation to two effects. First, aggregating a longer sequence of reward samples into a single relative-advantage estimate makes it harder to assign credit to individual actions, because the scalar return averages out the per-step contributions. Second, a longer rollout reduces the frequency of policy updates, so the policy is refined less often over the same number of environment steps. Both effects reduce the policy-update accuracy and thus degrade the achieved sum rate. This observation suggests $T=2$, which is suitable for the reward definition of our Dec-POMDP.

In Fig.~\ref{fig:compare_mn}, we show the training and testing curves for MAGRPO and the baselines under varying numbers of FAs per BS $M$ and numbers of BSs $N$. Three observations can be drawn. First, MAGRPO, MAPPO, and MAPPO-FA all converge in both training and testing across $M = 2, 3, 4, 5$ and $N = 1, 2, 3, 4$, which verifies the stability of the proposed critic-free training procedure. Second, the sum rate increases monotonically with $M$ in every configuration, confirming that additional FAs enlarge the feasible spatial-search region and allow each BS to discover channel conditions with higher SINR. Third, MAPPO-FA is consistently and clearly below MAGRPO and MAPPO, and the gap widens as $M$ grows. This directly isolates the benefit of jointly optimizing the FA positions, since MAPPO-FA differs only in that it freezes the FA positions to random values.

In Fig.~\ref{fig:performace_m} and \ref{fig:time_m}, we compare the testing sum rate and the training time of MAGRPO against the baselines under varying $M$, where the sum rates are obtained by applying an exponential moving average (EMA) with a smoothing factor of $0.99$ to the testing curves at the $8$-million-step checkpoint. Fig.~\ref{fig:performace_m} shows that MAGRPO achieves sum rates comparable to or slightly higher than MAPPO in most test cases, confirming that replacing the centralized critic and GAE with the group relative advantage does not compromise policy quality. This is expected, because both algorithms optimize the same clipped surrogate objective over the same on-policy data and differ only in how the advantage is estimated. Meanwhile, Figs.~\ref{fig:time_m} and \ref{fig:time_n} show that MAGRPO reduces the training time by about $20\%$ compared with MAPPO. We attribute this saving to the removal of the critic network, whose forward and backward passes are eliminated entirely. Finally, Fig.~\ref{fig:performace_m} shows that MAGRPO achieves $542.6\%$, $530.3\%$, $470.3\%$, and $1002.3\%$ gains over MAPPO-FA for $M = 2, 3, 4, 5$, respectively. The non-monotonic pattern of these gains is likely because the fixed random FA positions of MAPPO-FA can occasionally coincide with favorable channel conditions, whereas MAGRPO can always reposition the antennas; as $M$ grows, the freedom to jointly adjust all FA positions becomes increasingly valuable.

In Fig.~\ref{fig:performance_n} and \ref{fig:time_n}, we evaluate the testing sum rate and the training time versus the number of BSs $N$ with $M=6$, $K=2$, $G=16$, and $T=2$. Fig.~\ref{fig:performance_n} shows that the sum rates of MAGRPO, MAPPO, and MATD3 all grow with $N$, because a larger network serves more users and thus accumulates a higher network-wide sum rate. Importantly, MAGRPO matches or slightly exceeds MAPPO for every $N$, which indicates that the critic-free advantage estimation remains effective as the inter-cell interference becomes more severe. This robustness stems from the fact that the group relative advantage is computed from the actual network-wide returns, so it still reflects the global reward signal without requiring an explicit critic to model the growing state space. Moreover, Fig.~\ref{fig:time_n} shows that the training-time advantage of MAGRPO over MAPPO is preserved as $N$ grows, and Table~\ref{tab:runtime_reduction} indicates that the model-size and FLOPs reductions grow with $N$, because the removed critic network scales with the global-state dimension, which increases with $N$.

Table~\ref{tab:runtime_reduction} summarizes the training-time, model-size, and FLOPs reductions of MAGRPO over MAPPO and MATD3. The model-size and FLOPs reductions are substantial and grow with $N$; for example, with $M=6$ the model-size reduction over MAPPO increases from $47.4\%$ at $N=1$ to $69.4\%$ at $N=4$. This is because the critic network, which is removed in MAGRPO, dominates the total parameter count and its input dimension scales with $N$. In contrast, the training-time reduction is about $20\%$--$23\%$ and grows only mildly with $N$, since the actor forward and backward passes still dominate the per-step computation under parameter sharing and the MAPPO warm-up stage is included in the total training time. Over MATD3, which employs twin critics, MAGRPO achieves even larger reductions, reaching about $35\%$ in training time, $84.1\%$ in model size, and $84.2\%$ in FLOPs at $M=6$, $N=4$. These results confirm that the critic-free design of MAGRPO yields a lighter and faster model with essentially no loss in sum-rate performance.

\section{Conclusion}

This paper investigated the joint optimization of FA positions, beamforming, and transmit power in a multi-cell FA-assisted downlink network, which is a non-convex and high-dimensional stochastic optimization problem. We formulated the problem as a Dec-POMDP and proposed the critic-free MAGRPO under the CTDE paradigm. By replacing MAPPO's centralized critic and GAE with a group relative advantage derived from joint trajectory returns, MAGRPO eliminates the critic network while preserving decentralized execution, and reduces the per-step computational complexity by approximately half under parameter sharing.

Extensive simulations were conducted to evaluate MAGRPO against MAPPO, MATD3, and a fixed-position baseline (MAPPO-FA). The results show that MAGRPO achieves sum-rate performance higher than MATD3 and comparable to or slightly higher than MAPPO while reducing the training time by about $20\%$--$23\%$ compared with MAPPO and by about $35\%$ compared with MATD3, and it attains several-fold sum-rate gains over fixed-position configurations. Ablation studies further reveal that a sufficiently large group provides a more reliable relative-advantage baseline and smoother training, that a shorter trajectory length reduces the credit-assignment difficulty and improves the achieved sum rate, and that a suitable MAPPO warm-up stage balances the testing sum rate and the end-to-end training efficiency. These results demonstrate that MAGRPO accelerates MARL training without sacrificing the sum-rate performance, offering a promising and practical solution for FA-assisted wireless networks.

In future work, we will extend MAGRPO to more general settings, including imperfect CSI, user mobility, and larger-scale networks, and will further investigate its theoretical convergence and sample-efficiency properties.

\bibliographystyle{IEEEtran}

\begin{thebibliography}{30}
\providecommand{\url}[1]{#1}
\csname url@samestyle\endcsname
\providecommand{\newblock}{\relax}
\providecommand{\bibinfo}[2]{#2}
\providecommand{\BIBentrySTDinterwordspacing}{\spaceskip=0pt\relax}
\providecommand{\BIBentryALTinterwordstretchfactor}{4}
\providecommand{\BIBentryALTinterwordspacing}{\spaceskip=\fontdimen2\font plus
\BIBentryALTinterwordstretchfactor\fontdimen3\font minus
  \fontdimen4\font\relax}
\providecommand{\BIBforeignlanguage}[2]{{%
\expandafter\ifx\csname l@#1\endcsname\relax
\typeout{** WARNING: IEEEtran.bst: No hyphenation pattern has been}%
\typeout{** loaded for the language `#1'. Using the pattern for}%
\typeout{** the default language instead.}%
\else
\language=\csname l@#1\endcsname
\fi
#2}}
\providecommand{\BIBdecl}{\relax}
\BIBdecl

\bibitem{MIMO-1}
F. Tariq, M. R. A. Khandaker, K.-K. Wong, et al., ``A speculative study on 6G,'' \emph{IEEE Wireless Commun.}, vol. 27, no. 4, pp. 118–125, 2020.

\bibitem{MIMO-2}
R. Zhang, L. Cheng, S. Wang, et al., ``Integrated sensing and communication with massive MIMO: A unified tensor approach for channel and target parameter estimation,'' \emph{IEEE Trans. Wireless Commun.}, vol. 23, no. 8, pp. 8571–8587, 2024.

\bibitem{MIMO-3}
W. Wang and W. Zhang, ``Orthogonal projection-based channel estimation for multi-panel millimeter wave MIMO,'' \emph{IEEE Trans. Wireless Commun.}, vol. 68, no. 4, pp. 2173–2187, 2020.

\bibitem{FAS-Kit-1}
K.-K. Wong, A. Shojaeifard, et al., ``Fluid antenna systems,'' \emph{IEEE Trans. Wireless Commun.}, vol. 20, no. 3, pp. 1950–1962, 2021.

\bibitem{FAS-Kit-2}
K.-K. Wong and K.-F. Tong, ``Fluid antenna multiple access,'' \emph{IEEE Trans. Wireless Commun.}, vol. 21, no. 7, pp. 4801–4815, 2022.

\bibitem{FAS-Kit-3}
W. K. New, K.-K. Wong, H. Xu, et al., ``A tutorial on fluid antenna system for 6G networks: Encompassing communication theory, optimization methods and hardware designs,'' \emph{IEEE Commun. Surveys \& Tutorials}, vol. 27, no. 4, pp. 2325–2377, 2025.
\bibitem{FluidAntenna_Lu2025}
W.-J. Lu, C.-X. He, Y. Zhu, et al., ``Fluid antennas: Reshaping intrinsic properties for flexible radiation characteristics in intelligent wireless networks,'' \emph{IEEE Commun. Mag.}, vol. 63, no. 5, pp. 40-45, May 2025.
\bibitem{FluidAntenna_Hong2026}
H. Hong, K. K. Wong, H. Xu, et al., ``A contemporary survey on fluid antenna systems: Fundamentals and networking perspectives,'' \emph{IEEE Trans. Netw. Sci. Eng.}, vol. 13, pp. 2305-2328, 2026.
\bibitem{FluidAntenna_New2026}
W. K. New et al., ``Fluid antenna systems: Redefining reconfigurable wireless communications,'' \emph{IEEE J. Sel. Areas Commun.}, vol. 44, pp. 1013-1044, 2026, doi: 10.1109/JSAC.2025.3632097.
\bibitem{FluidAntenna_Wu2025}
T. Wu et al., ``Fluid antenna systems enabling 6G: Principles, applications, and research directions,'' \emph{IEEE Wireless Commun.}, doi: 10.1109/MWC.2025.3629597.
\bibitem{efrem2022transmit}
C. N. Efrem and I. Krikidis, ``Transmit and receive antenna port selection for channel capacity maximization in fluid-MIMO systems,'' \emph{IEEE Wireless Commun. Lett.}, vol. 13, no. 11, pp. 3202–3206, Nov. 2024.
\bibitem{Shuaixin}
S. Yang, Y. Xiao, Y. L. Guan, et al., ``BER performance optimization for fluid antenna-aided wireless communications,'' \emph{IEEE J. Sel. Areas Commun.}, vol. 44, pp. 1177–1192, 2026.
\bibitem{tang2025capacity}
B. Tang, H. Xu, K. -K. Wong, et al., ``Capacity maximization of uplink with fluid antenna system at both ends,'' \emph{IEEE Trans. Wireless Commun.}, vol. 24, no. 12, pp. 10578–10593, Dec. 2025.

\bibitem{Tianyi}
T. Liao, W. Guo, H. He, et al., ``Joint beamforming and antenna position optimization for fluid antenna-assisted MU-MIMO networks,'' \emph{IEEE J. Sel. Areas Commun.}, vol. 44, pp. 1209–1226, 2026.
\bibitem{Chen2026}
Y. Chen, B. Xu, S. Li, et al., ``Analysis and optimization for low-latency communications in slow fluid antenna multiple access systems,'' \emph{IEEE J. Sel. Areas Commun.}, vol. 44, no. 4, pp. 1290–1306, 2026.

\bibitem{ISAC}
Q. Zhang, M. Shao, T. Zhang, et al., ``An efficient sum-rate maximization algorithm for fluid antenna-assisted ISAC system,'' \emph{IEEE Commun. Lett.}, vol. 29, no. 1, pp. 200–204, 2025.
\bibitem{Ho2025}
M. C. Ho, T. D. T. Tam, T. S. Do, and S. Cho, ``Proximal policy optimization for latency minimization in FL-assisted fluid antenna systems with MC-NOMA,'' in \emph{Proc. 2025 30th Asia-Pacific Conf. Commun. (APCC)}, pp. 1–6, 2025.

\bibitem{zhang2026indoor}
T. Zhang, Q. Li, S. Wang, et al., ``Indoor fluid antenna systems enabled by layout-specific modeling and group relative policy optimization,'' \emph{IEEE Trans. Wireless Commun.}, vol. 25, pp. 9312–9330, Jan. 2026.

\bibitem{waqar2024opportunistic}
N. Waqar, K.-K. Wong, C.-B. Chae, et al., ``Opportunistic fluid antenna multiple access via team-inspired reinforcement learning,'' \emph{IEEE Trans. Wireless Commun.}, vol. 23, pp. 12068–12083, Sept. 2024.
\bibitem{weng2024learning}
C. Weng, Y. Chen, L. Zhu, and Y. Wang, ``Learning-based joint beamforming and antenna movement design for movable antenna systems,'' \emph{IEEE Wireless Commun. Lett.}, vol. 13, pp. 2120–2124, Aug. 2024.

%\bibitem{amiri2025movable}
%M. Amiri, A. Mohammadzadeh, F. Zeinali, et al., ``Movable antenna SWIPT systems with STAR-RIS: A meta DRL approach,'' \emph{IEEE Trans. Wireless Commun.}, vol. 24, pp. 6362–6375, Aug. 2025.

\bibitem{wang2024fluid}
C. Wang, G. Li, H. Zhang, et al., ``Fluid antenna system liberating multiuser MIMO for ISAC via deep reinforcement learning,'' \emph{IEEE Trans. Wireless Commun.}, vol. 23, pp. 10879–10894, Sept. 2024.


\bibitem{wei2025movable}
H. Wei, W. Wang, W. Ni, et al., ``Movable-antenna enabled cell-free networks,'' \emph{IEEE Trans. Veh. Technol.}, vol. 74, no. 10, pp. 16533–16537, 2025.

\bibitem{zhu2025joint}
J. Zhu, L. Feng, X. Wang, et al., ``Joint beamforming, user association, and antenna position optimization in movable antenna-assisted cell-free massive MIMO,'' \emph{IEEE Trans. Netw. Sci. Eng.}, vol. 13, pp. 4153–4169, 2026.

\bibitem{Zhao2026FA}
J. Zhao, M. Chen, Z. Yang, H. Xu, C. Pan, T. Q. S. Quek, and K.-K. Wong, ``Delay efficient FA-assisted satellite communication network with mobile edge computing,'' \emph{IEEE Internet of Things Journal}, vol. 13, no. 5, pp. 9435--9449, Mar. 2026.

\bibitem{mao2025joint}
T. Mao, Z. Chu, Y. Wang, Z. Zhu, W. Hao, D. Mi, and C. Pan, ``Joint time scheduling and port activation design for fluid antenna-empowered wireless powered communication networks,'' \emph{IEEE Internet Things J.}, vol. 12, no. 13, pp. 22904–22914, Jul. 2025.

\bibitem{wu2026fluid}
N. Wu, L. Wang, R. Jiang, B. Li, Y. Feng, and A. Nallanathan, ``Fluid antennas and stacked intelligent metasurfaces enabled multi-user ISAC for low-altitude wireless networks,'' \emph{IEEE Trans. Veh. Technol.}, vol. 75, no. 7, pp. 1–17, Jul. 2026.

\bibitem{Ghadi2025}
F. R. Ghadi, et al., ``Fluid antenna multiple access with simultaneous non-unique decoding in strong interference channel,'' \emph{IEEE Trans. Wireless Commun.}, vol. 24, no. 12, pp. 10183–10195, Dec. 2025.
\bibitem{Li2026}
X. Li, Q. Cui, B. Zhao, et al., ``SWIPT optimization design for multi-RIS-aided cell-free IoT networks with fluid antenna,'' \emph{IEEE Trans. Wireless Commun.}, vol. 25, pp. 10484–10497, 2026.
\bibitem{Li2025MATD3}
Q. Li, W. Wang, Y. Li, F. Yu, C. Zhang, and Y. Huang, ``Deep reinforcement learning for movable antenna-assisted cell-free networks,'' \emph{IEEE Wireless Commun. Lett.}, vol. 14, pp. 2783–2787, Sept. 2025.






\bibitem{su2024cd}
C. Su, R. Wu, Y. Zhu, and Q. Hu, ``CD-MAPPO: Centralized-decentralized multi-agent proximal policy optimization in multi-cell networks,'' in \emph{Proc. IEEE/CIC Int. Conf. Commun. China (ICCC) Workshops}, pp. 1–6, 2024.
\bibitem{CTDE}
R. Lowe, Y. I. Wu, A. Tamar, et al., ``Multi-agent actor-critic for mixed cooperative-competitive environments,'' in \emph{Proc. Advances in Neural Information Processing Systems (NeurIPS)}, vol. 30, 2017.
\bibitem{MAPPO}
C. Yu, A. Velu, E. Vinitsky, et al., ``The surprising effectiveness of PPO in cooperative multi-agent games,'' in \emph{Proc. Advances in Neural Information Processing Systems (NeurIPS)}, vol. 35, pp. 24611–24624, 2022.
\bibitem{GRPO}
Z. Shao, P. Wang, Q. Zhu, et al., ``Deepseekmath: Pushing the limits of mathematical reasoning in open language models,'' \emph{arXiv preprint arXiv:2402.03300}, 2024.
\bibitem{kingma2014adam}
D. P. Kingma and J. Ba, ``ADAM: A method for stochastic optimization,'' \emph{arXiv preprint arXiv:1412.6980}, 2014.
\bibitem{PS}
H. Qin, Z. Liu, C. Lin, et al., ``GRADPS: Resolving futile neurons in parameter sharing network for multi-agent reinforcement learning,'' in \emph{Proc. International Conference on Machine Learning (ICML)}, 2025.



\bibitem{Ackermann2019}
J. Ackermann, V. Gabler, T. Osa, and M. Sugiyama, ``Reducing overestimation bias in multi-agent domains using double centralized critics,'' in \emph{Proc. Advances in Neural Information Processing Systems (NeurIPS) Deep RL Workshop}, Vancouver, Canada, Dec. 2019.

\bibitem{zhang2025multi}
Y. Zhang and D. Guo, ``Multi-agent reinforcement learning for multi-cell spectrum and power allocation,'' \emph{IEEE Trans. Commun.}, vol. 73, no. 8, pp. 5980–5992, Aug. 2025.


\end{thebibliography}

\end{document}